\documentclass[preprint,review,3p,times,sort&compress]{elsarticle}
\usepackage{import}
\usepackage{graphics}
\usepackage{arydshln}
\usepackage{graphicx}
\usepackage{subfigure}
\usepackage[numbers]{natbib}
\usepackage{psfrag}
\usepackage{dirtytalk}
\usepackage{dutchcal}
\usepackage{xfrac}
\usepackage{amssymb}
\usepackage{amsmath}
\usepackage{mathrsfs}
\usepackage{caption}
\usepackage{amsthm}
\usepackage{xcolor}
\usepackage{rotating}
\usepackage{lscape}
\usepackage{pdflscape}
\usepackage{booktabs}
\usepackage{multirow}
\usepackage{hyperref}
\usepackage{lineno}
\usepackage{float}
\usepackage{epstopdf}
\usepackage{changes}
\usepackage{lipsum}
\usepackage{tikz}

\usepackage{bigints}

\AtBeginDocument{\let\latexlabel\label}
\usepackage{cancel}

\begin{document}
\begin{frontmatter}
\title{\vspace{-6cm} \textbf{Acoustic absorption and generation in ducts of smoothly varying area sustaining a mean flow and a mean temperature gradient}}
\date{\today}
%
\author{Saikumar R.~Yeddula \corref{cor1}}
\ead{s.yeddula18@imperial.ac.uk}

\author{Renaud Gaudron \corref{cor2}}
\ead{r.gaudron@imperial.ac.uk}

\author{Aimee S.~Morgans \corref{cor3}}
\ead{a.morgans@imperial.ac.uk}

\address{Department of Mechanical Engineering, Imperial College London, London, UK}
\cortext[cor1]{Corresponding author.}
\begin{abstract}
In ducts with varying cross-sectional area and sustaining a subsonic non-isentropic mean flow, the axially varying flow conditions affect the acoustic energy balance of the system. This is significant in understanding and controlling thermo-acoustic phenomena, particularly in combustors. This work aims at quantifying the acoustic energy change in such configurations, using the acoustic absorption coefficient, $\Delta$. The acoustic response of the duct to acoustic forcing is determined using an analytical model, neglecting the effect of entropy fluctuations on the acoustic field, and subsequently, $\Delta$ is estimated. The model predictions of $\Delta$ are validated using a linearised Euler equations (LEEs) solver. The model was found to be accurate for Mach numbers below $0.25$, provided the lower frequency limit set by the analytical solution is satisfied. For conically varying area ducts with linear mean temperature gradient, it was observed that $\Delta$ showed very little dependence on frequency, and that the absolute value of $\Delta$ tended to be maximised when the upstream boundary was anechoic rather than non-anechoic. More importantly, $\Delta$ was also observed to show stronger dependence on the mean temperature gradient than area gradient variation for such configurations. Further parametric and optimisation studies for $\Delta$ revealed a crucial finding that a positive mean temperature gradient, representing a heated duct caused acoustic energy absorption. Similarly, a negative mean temperature gradient, representing a cooled duct caused acoustic energy generation -- a key result of this analysis. This behaviour was shown to be consistent with a simplified analysis of the acoustic energy balance. Based on this finding, a linearly proportional reduction in acoustic energy generation was achieved by changing the mean temperature gradient.
\end{abstract}
\end{frontmatter}
\section{Background}\label{sec:Background}
Thermoacoustic instabilities are a major problem in boilers, low NO$_x$ gas-turbine combustors, and rocket engine combustors,  \citep{Liu_JP&P_2014}. They occur due to a positive feedback mechanism between the unsteady flame and the acoustic waves, that lead to self-sustained oscillations \citep{Dowling_JP&P_2003, Lieuwen_2005}. These oscillations increase the overall noise emissions of the engine, lead the system to operate in off-design conditions, and can also result in premature component failure or permanent structural damage \citep{Lieuwen_2005}. Understanding the acoustic energy balance for given operating conditions can help in identifying the regions susceptible to instability, during the preliminary design phase. However, flow-acoustic coupling in such combustors is often complex owing to variations of both the duct cross-sectional area, the stream-wise mean temperature, in the presence of large subsonic mean flow velocities \citep{Dokumaci_JSV_1998a}. Acoustic behaviour in such duct configurations is also relevant to automotive exhausts, heat exchangers, and other thermo-fluid flow devices. Full numerical simulations aimed at characterising the mean and fluctuating variables concomitantly are computationally costly, due to the disparity in the time and length scales. 

Computational low-order acoustic network models have gained much popularity, as the equations for the mean flow and the acoustics can be solved sequentially \citep{Heckl_A&HA_1986, Candel_ISC_1992, Dowling_JSV_1995, Keller_ZAMP_1985, Keller_AIAA_1995, Candel_PCI_2002, Dowling_JP&P_2003, Han_CNF_2015Oct, Laera_CNF_2017, Xia_CST_2019, Schaefer_AIAA_2019}. As any parametric analysis to identify stability boundaries requires consideration of many cases; analytical solutions that can characterize the acoustic behaviour for any given geometry, mean flow and acoustic boundary conditions offer significant advantages \citep{Marble_JSV_1977, Easwaran_JSV_1992, Subrahmanyam_JSV_2001, Tyagi_JFM_2003, Tyagi_JFM_2005, Duran_JFM_2013}. Unlike the analytical solutions, certain solutions require simple numerical computations and are termed as ``semi-analytical" solutions. Analytical (respectively semi-analytical) solutions are grid independent (respectively convergent) and are computationally fast. They also provide more physical insight into the acoustic behaviour across various mean flow and boundary conditions.

As the modal frequencies at which thermoacoustic phenomena occur in aviation and power generation combustors are typically low, the effects of thermal and molecular diffusion are negligible. Hence, the linearised Euler equations (LEEs) are accurate enough to describe the acoustic behaviour. 
Several high frequency approximation solutions, based on the WKB method \citep{Schlissel_HM_1977} have been proposed for solving the LEEs. Cummings \citep{Cummings_JSV_1977} used the WKB method and presented a semi-analytical solution for the acoustic field in a uniform area duct sustaining a temperature gradient at very low Mach number and no entropy perturbations. Dokumaci \cite{Dokumaci_JSV_1998a} extended the analysis to the case of a varying area duct carrying isentropic non-zero mean flow. Using an adaptive WKB method, Li \& Morgans~\citep{Li_JSV_2017} analysed the acoustic field in straight ducts sustaining practical non-isentropic mean flows with a temperature gradient. Rani and Rani \citep{Rani_JSV_2018} presents approximate (WKB and WKB2) solutions to the acoustic field in quasi 1-D varying area ducts with isentropic flow. A semi-analytical solution for the LEEs was proposed in our previous study \citep{Yeddula_JSV_2020}, for varying area ducts with mean flow, where the mean flow does not necessarily have to be isentropic. Further assuming low flow Mach numbers, analytical solutions were also obtained for certain area and mean temperature profiles but no insight into the acoustic energy generation or absorption in these ducts was reported. These solutions are used in this work for the estimation of the acoustic field to further understand the acoustic energy balance in duct flows. 

The interaction between the acoustics and the flow in terms of energy balance was first investigated by Cantrell \& Hart \citep{Cantrell_JASA_1964} for irrotational, isentropic, subsonic mean flows. Formal definitions for the acoustic energy flux and acoustic energy density were also first presented in their work. The acoustic energy balance was later extended to non-uniform flows by Morfey \citep{Morfey_JSV_1971}. The model was further improved to include the effects of mean and unsteady heat addition by Bloxsidge \textit{et al.} \citep{Bloxsidge_AIAA_1988}, for one-dimensional flow. For arbitrary steady flows, a general disturbance energy corollary was first proposed by Myers \citep{Myers_JFM_1991}, which included both acoustic and non-acoustic disturbance energy terms. Giauque \textit{et al.} \citep{Giauque_CTR_2006, Brear_JFM_2012} further extended Myers energy corollary to flows with gaseous combustion, by including species and heat release terms. For straight ducts with mean and fluctuating heat release, Karimi \textit{et al.} \citep{Karimi_JFM_2008} analysed the acoustic and disturbance energy contributions. A frequency limit, called the `entropic corner frequency' was proposed, above which the effect of acoustic-entropy coupling is negligible. At frequencies larger than this limit, the total disturbance energy can be approximated as acoustic energy \citep{Karimi_JFM_2008}. 

Although acoustic energy balance studies exist, they are generalised and do not explicitly consider duct flows with area and mean temperature variations. Characterisation of the energy balance with minimal parameters reduces the computational effort in analysing many practical duct configurations, like combustors and heat exchangers. 

The generation or dissipation of acoustic energy in ducts can be fully characterised by a single parameter, $\Delta$, called the acoustic absorption coefficient \citep{Morfey_JSV_1971}. It quantifies the loss of the acoustic energy flux across the duct, for instance $\Delta >0$ (respectively $\Delta < 0$) implies acoustic energy absorption (respectively generation), while $\Delta = 0$ corresponds to no change in the acoustic energy, inside the duct. This characterisation of the acoustic energy balance for duct flows, in terms of $\Delta$ was first investigated by Gaudron \textit{et al.} \citep{Gaudron_ICSV_2019, Gaudron_GT&P_2021}. The final expression for $\Delta$ was shown to be a function of only a few mean flow parameters and the wave amplitudes at the upstream and downstream ends of the duct. The acoustic wave amplitudes can be obtained using numerical simulations, experiments or analytical models.

The primary objective of the present work is to develop a simplified analytical framework to estimate the acoustic absorption coefficient $\Delta$, in ducts with (i) varying cross-sectional area (ii) varying stream-wise mean temperature and (iii) low Mach number subsonic mean flow, which does not have to be isentropic. This is achieved by characterising the acoustic flow field variables using the analytical solution proposed in \citep{Yeddula_JSV_2020} and coupling it with the approach proposed in \citep{Gaudron_ICSV_2019} for estimating $\Delta$. This allows the exposure of direct linkage between duct heating/cooling and acoustic damping/generation, and it is one of the main contributions of this paper. This linkage is also investigated by utilising both the acoustic energy balance analysis, in line with \citep{Morfey_JSV_1971}, and a simplified mathematical analysis for a constant area duct with low Mach number flow.

The WKB solution used in this analysis requires the frequency to be large in some sense \citep{Yeddula_JSV_2020}, while also below the cut-on frequency of the duct, such that the acoustic field remains one-dimensional. This solution also assumes the effect of the entropy fluctuations on the acoustic field to be negligible. This is a valid approximation because at the considered high frequency regimes, the entropy waves are strongly attenuated due to shear dispersion in practical flow scenarios \citep{Morgans_JFM_2013, Xia_FT&C_2018, Dowling_JPE_2000, Sattelmayer_GT&P_2003, Giusti_AIAA_2017}. Also, at the high frequencies considered, the total disturbance energy can be approximated by the acoustic energy \citep{Karimi_JFM_2008}. Further, as the flow Mach numbers in combustors are also maintained low ($M \sim 0.2$) for a stable flame, the effect of the entropy fluctuations on the acoustic field is negligible \citep{Dokumaci_JSV_1998b, Li_IJSCD_2018}.

This work is organised as follows. After introducing the problem formulation, a numerical mechanism for solving the linearised Euler equations (LEEs) \citep{Dokumaci_JSV_1998b} and computing $\Delta$ for non-isentropic flows, is introduced in Section \ref{sec:ProblemFormulation}. In Section \ref{sec:Analytical}, the wave amplitudes are estimated using the semi-analytical solution in \citep{Yeddula_JSV_2020}, and an explicit expression for $\Delta$ is determined. Section \ref{sec:Setting the validation test case} describes the validation case in which comparison of the numerical evaluation with the analytical estimate is performed for a conical duct with linearly varying stream-wise mean temperature. For this particular configuration, the analytical solution of the acoustic field presented in \citep{Yeddula_JSV_2020}, which depends only on the local flow properties and frequency, is used. The validation results are presented in Section \ref{sec:Results}, which constitutes a parametric analysis followed by an optimization search which identifies extrema for $\Delta$ across different linear area and mean temperature gradients, flow Mach numbers, frequencies and acoustic boundary conditions. In Section \ref{sec:EnergyAnalysis}, an energy analysis is performed to understand the acoustic behaviour in ducts with mean temperature gradient. Conclusions are drawn in the final Section \ref{sec:Conclusions}. 

\begin{figure}[tbp!] 
\def\svgwidth{0.8\textwidth}
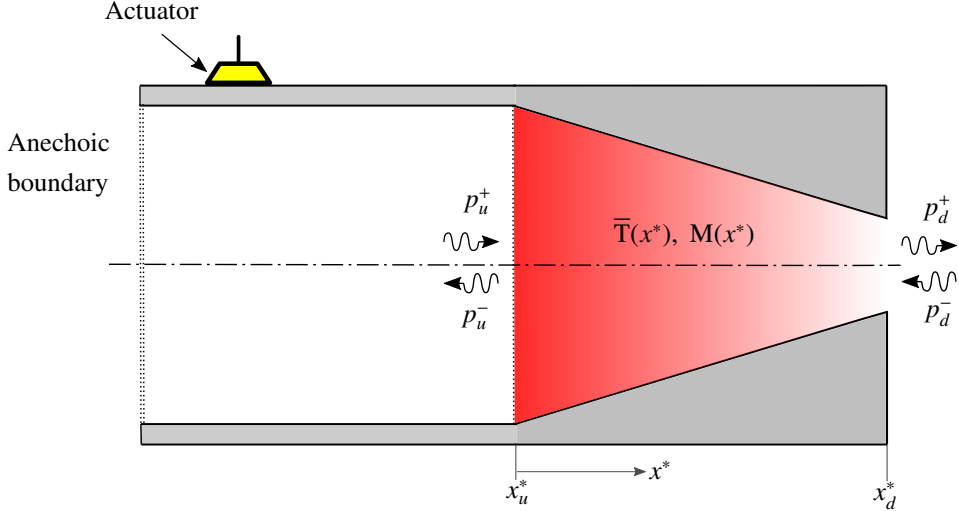
\centering
\caption{Varying area duct sustaining stream-wise mean temperature gradient and subsonic mean flow. $p^{\pm}$ represent the downstream and upstream propagating acoustic wave amplitudes. Subscript `u, d' denotes the upstream and downstream ends of the duct.}
\label{fig:Figure1}
\end{figure}

\section{Problem formulation}\label{sec:ProblemFormulation}
 The mean flow in varying area ducts, an example of which is shown in Fig.~\ref{fig:Figure1}, can be considered to be quasi one-dimensional, if the variations in the radial direction associated with boundary layer growth are negligible. The conservation equations for mass, momentum, and energy, in such quasi 1-D perfect inviscid flow are as follows:
        \begin{equation*}
                  \refstepcounter{equation}\latexlabel{eq:massCons}
                  \refstepcounter{equation}\latexlabel{eq:momCons}
                  \refstepcounter{equation}\latexlabel{eq:enrCons}
                  \refstepcounter{equation}\latexlabel{eq:EqofState}
                       A \dfrac{ \partial \rho  }{ \partial t }  + \dfrac{\partial(\rho Au)}{\partial x} = 0; \quad
                            \dfrac{ \partial u }{ \partial t}+u\dfrac{\partial{u}}{\partial x} + \dfrac{1}{\rho}\dfrac{\partial p}{\partial x} = 0;\quad
                      \dfrac{ \partial {s} }{ \partial t } +{u}\dfrac{\partial{s}}{\partial x} = \dfrac{{R}_{g}}{p}{\dot{Q}} \quad \text{with} \quad {p} = {\rho} {R}_{g} {T},
                 \tag{\ref*{eq:massCons}, \ref*{eq:momCons}, \ref*{eq:enrCons}, \ref*{eq:EqofState}}
       \end{equation*}
 where ${\rho}$, ${p}$, ${T}$, ${u}$, ${s}$ denote density, pressure, temperature, axial velocity and entropy and are all functions of spatial and temporal coordinates $x$, $t$, while the cross-sectional area $A$ is only a function of $x$. ${R}_{g}$ (the gas constant) and $\gamma$ (the ratio of the specific heats) are considered constant, with values of $287 \text{J} \text{kg}^{-1} \text{K}^{-1}$ and $1.4$ respectively. 

The linearisation principle allows the flow variables to be written as the sum of mean time-averaged and fluctuating time-dependant components, denoted by $\overline{(\;\:)}$ and $(\;)'$ respectively. This gives,
\begin{equation}\label{eq:LinearisingFlowVariables}
\rho(x,\:t) = \overline{\rho}(x) + \rho'(x,\:t); \;\; T(x,\:t) = \overline{T}(x) + T'(x,\:t); \;\; p(x,\:t) = \overline{p}(x) + p'(x,\:t);\;\; u(x,\:t) = \overline{u}(x) + u'(x,\:t);\;\; s = \overline{s}(x) + s'(x,\:t).
\end{equation}
The variations in area and stream-wise mean temperature are represented by non-dimensional parameters, defined as,  

\begin{equation*}
     \refstepcounter{equation}\latexlabel{eq:Alphadef}
     \refstepcounter{equation}\latexlabel{eq:Betadef}
     \alpha^* = \dfrac{1}{A}\dfrac{\text{d}A}{\text{d}x^*}, \quad \beta^* = \dfrac{1}{\overline{T}}\dfrac{\text{d}\overline{T}}{\text{d}x^*}, \;\textrm{where}, \; x^* = \dfrac{x}{L},
     \tag{\ref*{eq:Alphadef} - \ref*{eq:Betadef}}\label{eq:AlphaBetaDefinition}
\end{equation*}   
and $L$ is the duct length. It is assumed that there are no heat-fluctuations $\dot{Q}'(x^*\:,t) = 0$ in this analysis. Using thermodynamic relations, the mean heat transfer per unit volume from the duct, $\overline{\dot{Q}}$, is then given by,
\begin{equation}
     \overline{\dot{Q}} = \dfrac{\overline{\rho}\:\overline{u}}{L}\left(\dfrac{\gamma R_g}{\gamma - 1}\dfrac{\text{d}\overline{T}}{\text{d}x^*}+\overline{u}\dfrac{\text{d}\overline{u}}{\text{d}x^*}\right).
\label{eq:HeatTransferDefinition}
\end{equation}   
The steady components of the conservation equations (Eqs.~\eqref{eq:massCons} -~\eqref{eq:enrCons}) allows us to further write the mean flow gradients in terms of the non-dimensional area and mean temperature gradients, $\alpha^*$ and $\beta^*$, respectively, as follows,
\begin{equation*}
                  \refstepcounter{equation}\latexlabel{eq:graduNonIsentropic}
                  \refstepcounter{equation}\latexlabel{eq:gradpNonIsentropic}
                  \refstepcounter{equation}\latexlabel{eq:gradrhoNonIsentropic}
                  \refstepcounter{equation}\latexlabel{eq:gradcNonIsentropic}
                       \dfrac{1}{\overline{u}}\dfrac{\text{d}\overline{u}}{\text{d}x^*} = \dfrac{\beta^*-\alpha^*}{1 - \gamma{M}^{2}}; \quad     \dfrac{1}{\overline{p}}\dfrac{\text{d}\overline{p}}{\text{d}x^*} = \dfrac{-\gamma{M}^{2}\left(\beta^* - \alpha^* \right)}{1 - \gamma{M}^{2}}; \quad     \dfrac{1}{\overline{\rho}}\dfrac{\text{d}\overline{\rho}}{\text{d}x^*} = \dfrac{-\left(\beta^* - \alpha^* \gamma {M}^{2}\right)}{1 - \gamma{M}^{2}};\quad      \dfrac{1}{\overline{c}}\dfrac{\text{d}\overline{c}}{\text{d}x^*} = \dfrac{\beta^*}{2},
                 \tag{\ref*{eq:graduNonIsentropic} - \ref*{eq:gradcNonIsentropic}}
\end{equation*}
where $M$ is the local Mach number of the flow, $M\:=\:\dfrac{\overline{u}}{\overline{c}}$.

 Similarly, the linearised form of the Euler equations obtained in the time domain are given by equating the time-dependent fluctuating quantities of the conservation equations (Eqs.~\eqref{eq:massCons} -~\eqref{eq:enrCons}) (refer to \ref{sec:appendixa} for the full derivation). This results in:
\begin{equation}\label{eq:NormalisedMassEnergy}
    \dfrac{ \partial   }{ \partial t }\left[\dfrac{{p'}}{\gamma \overline{p}}\right] + \dfrac{\overline{u}}{L} \dfrac{\partial}{\partial x^*}\left(\dfrac{{p'}}{\gamma \overline{p}} + \dfrac{{u'}}{\overline{u}} \right) +  \left(\dfrac{\gamma-1}{\gamma \overline{p}}\right)\overline{\dot{Q}}\left(\dfrac{{p'}}{\overline{p}} + \dfrac{{u'}}{\overline{u}}\right) = 0,
\end{equation}
\begin{equation}\label{eq:NormalisedMomentum}
    \dfrac{ \partial }{ \partial t }\left[\dfrac{{u'}}{\overline{u}}\right] + \dfrac{\overline{u}}{L} \dfrac{\partial}{\partial x^*}\left[\dfrac{{u'}}{\overline{u}}\right] +   \dfrac{1}{L} \dfrac{\textrm{d}{\overline{u}}}{\textrm{d} x^*}\left[\left(\gamma - 1\right)\dfrac{{p'}}{\gamma \overline{p}}+ \dfrac{{u'}}{\overline{u}} - \sigma'\right] + \dfrac{\overline{c}^2}{\overline{u} L}\dfrac{\partial}{\partial x^*} \left[\dfrac{{p'}}{\gamma \overline{p}}\right] = 0,
\end{equation}
\begin{equation}\label{eq:NormalisedEnergy}
    \dfrac{ \partial {\sigma'}}{ \partial t } + \dfrac{\overline{u}}{L} \dfrac{\partial \sigma'}{\partial x^*} =  \dfrac{R_g \overline{\dot{Q}}}{\overline{p}},
\end{equation}
where, ${\sigma}' = \dfrac{{s}'\left(\gamma - 1\right)}{\gamma R_g}$. By assuming a harmonic time-dependence for the fluctuating quantities, $y' = \hat{y}{\text{e}}^{\text{i}\omega t}$, where, $\omega$ is the complex angular frequency and $\text{i} = \sqrt {-1}$, the above equations are written in the frequency domain. The final equations thus read,
\begin{equation}\label{eq:LEE1NIfr}
    \begin{aligned}
\left(\textrm{i}\omega L + {\gamma}^{2}{M}^{2}\dfrac{\text{d}\overline{u}}{\text{d}x^*} + \gamma \overline{u} \dfrac{1}{\overline{T}}\dfrac{\text{d}\overline{T}}{\text{d}x^*}\right)\hat{p} + \overline{u}\dfrac{\text{d}\hat{p}}{\text{d}x^*} +
 \left(\dfrac{\text{d}\overline{p}}{\text{d}x^*}\left(\dfrac{1}{{M}^{2}} + 1 - \gamma\right) + \gamma\overline{p} \dfrac{1}{\overline{T}}\dfrac{\text{d}\overline{T}}{\text{d}x^*}\right)\hat{u} + \gamma \overline{p} \dfrac{\text{d}\hat{u}}{\text{d}x^*} = 0   
    \end{aligned}
\end{equation}
\begin{equation}\label{eq:LEE2NIfr}
    \left(\textrm{i}\omega L + \dfrac{\text{d}\overline{u}}{\text{d}x^*}\right)\hat{u} +
    \overline{u}\dfrac{\text{d}\hat{u}}{\text{d}x^*} + \overline{u}\dfrac{\text{d}\overline{u}}{\text{d}x^*}\dfrac{\hat{p}}{\gamma \overline{p}} + \dfrac{1}{\overline{\rho}}\dfrac{\text{d}\hat{p}}{\text{d}x^*} = \overline{u}\dfrac{\text{d}\overline{u}}{\text{d}x^*}\hat{\sigma}
\end{equation}
\begin{equation}\label{eq:LEE3NIfr}
    \textrm{i}\omega L \: \hat{\sigma}+\overline{u}\dfrac{\text{d} \hat{\sigma}}{\text{d}x^*} = \overline{u}\left(\dfrac{\alpha^* {M}^{2}\left(\gamma-1\right)-\beta^* \left(1-{M}^{2}\right)}{1-\gamma{M}^{2}}\right) \left(\dfrac{\hat{u}}{\overline{u}} + \dfrac{\hat{p}}{\overline{p}}\right).
\end{equation}

On dividing Eq. \eqref{eq:LEE1NIfr} by $\overline{u}$, and multiplying Eq. \eqref{eq:LEE2NIfr} by $\overline{\rho}$, coefficients of ${\text{d}\hat{p}}/{\text{d}x^*}$ are normalized, and the equations become, 
\begin{equation*}
                  \refstepcounter{equation}\latexlabel{eq:NormLEE1}
                  \refstepcounter{equation}\latexlabel{eq:NormLEE2}
                   A\hat{p} + \dfrac{\text{d}\hat{p}}{\text{d}x^*} + B\hat{u} + C\dfrac{\text{d}\hat{u}}{\text{d}x^*} = 0; \qquad D\hat{p} + \dfrac{\text{d}\hat{p}}{\text{d}x^*} + E\hat{u} + F\dfrac{\text{d}\hat{u}}{\text{d}x^*} = I \hat{\sigma},
                 \tag{\ref*{eq:NormLEE1}, \ref*{eq:NormLEE2}}
\end{equation*}
where the coefficients $A, B, C, D, E, F$, are functions of $x^*$ and frequency in terms of the Helmholtz number, He$ = \omega \: L / \overline{c}$, and have the following expressions,
\begin{equation}
                       A = \left(\dfrac{\text{i} \text{He}}{M} + {\gamma}^{2} {M}^{2}\dfrac{1}{\overline{u}} \dfrac{\text{d}\overline{u}}{\text{d}x^*} + \gamma \beta^*  \right), \;
                       B = \dfrac{\overline{\rho}\: \overline{c} }{M} \dfrac{1}{\gamma \overline{p}}  \left(\dfrac{\text{d}\overline{p}}{\text{d}x^*}\left(\dfrac{1}{M^2} + 1 - \gamma\right) + \beta^* \right),\;
                       C = \dfrac{\overline{\rho}\: \overline{c} }{M},\;
                         \refstepcounter{equation}\tag{\theequation a-c}\label{eq:CoefficientsLEE}
\end{equation}
\begin{equation}
 D = {M^2}\dfrac{1}{\overline{u}}\dfrac{\text{d}\overline{u}}{\text{d}x^*}, \;
                       E =  {\overline{\rho}\: \overline{c}} \left({\text{i} \text{He}} + M\dfrac{1}{\overline{u}}\dfrac{\text{d}\overline{u}}{\text{d}x^*}\right),\; 
                       F = \overline{\rho}\: \overline{c} M, \;
                       I = \overline{\rho} \: \overline{u} \dfrac{\text{d}\overline{u}}{\text{d}x^*}.
                         \refstepcounter{equation}\tag{\theequation a-d}\label{eq:Coefficients2LEE}                       
\end{equation}

On further expressing the pressure and velocity fluctuations in terms of the amplitudes of the acoustic waves propagating in the downstream (+) and upstream (-) direction \citep{Dokumaci_JSV_1998a}, using,
\begin{equation}\label{eq:phatuhatDefinitionDokumaci}
 {\hat{p}\left(x^*,\omega\right)}= {{{p}}^+ + {p}^-},\quad {\hat{u}\left(x^*,\omega\right)} =  \dfrac{1}{\overline{\rho} \: \overline{c}}\left({{p}^+ - {p}^-}\right),\;
\end{equation}
and setting $\hat{\epsilon} = \gamma \overline{p} \hat{\sigma}$, the linearised equations are recast into a first-order system of differential equations for wave amplitudes of acoustic ($p^{\pm}$) and entropy ($\hat{\epsilon}$) components, given by,
\begin{equation}\label{eq:dpplusminus}
\begin{aligned}
 \dfrac{\text{d}{p}^{\pm}}{\text{d}x^*} & = {p}^{+}  \overbrace{\left[\dfrac{-1}{2\left(1 \pm M\right)} \left(\left(D \pm A M\right) + \dfrac{1}{\overline{\rho} \:\overline{c}} \left(E \pm B M \right) + \dfrac{\text{d}}{\text{d}x}\left(\dfrac{1}{\overline{\rho} \: \overline{c}}\right) \left(F \pm C M\right)\right)\right]} ^ {{C}_{11} \;(+) \; \text{or} \; {C}_{21}\;(-)}   \\ & + {p}^{-}  \overbrace{\left[\dfrac{-1}{2\left(1 \pm M\right)} \left(\left(D \pm A M\right) - \dfrac{1}{\overline{\rho} \overline{c}} \left(E \pm B M \right) - \dfrac{\text{d}}{\text{d}x}\left(\dfrac{1}{\overline{\rho} \overline{c}}\right) \left(F \pm C M\right)\right)\right]} ^ {{C}_{12}  \;(+) \; \text{or} \; {C}_{22}\; (-) } + \:\hat{\epsilon} \overbrace{ \left[\dfrac{ \:M^2}{2\left(1 \pm M\right)}\dfrac{\beta^* - \alpha^*}{1 - \gamma M^2}\right]} ^ {{C}_{13} \;(+)\; \text{or} \; {C}_{23} \; (-)},
\end{aligned}
\end{equation}
\begin{equation}\label{eq:dsigma}
\begin{aligned}
\dfrac{\text{d}\hat{\epsilon}}{\text{d}x} = {{p}^{+}} \overbrace{\left[\zeta \left(\gamma + \dfrac{1}{M}\right)\right]} ^ {{C}_{31}}  + \; {{p}^{-}} \overbrace{\left[\zeta \left(\gamma - \dfrac{1}{M}\right)\right]} ^ {{C}_{32}}  - \hat{\epsilon} \overbrace{\left[{\dfrac{\text{i} \text{He}}{M}} - \dfrac{1}{\overline{p}}\dfrac{\text{d}\overline{p}}{\text{d}x^*}\right]} ^ {{C}_{33}};\quad \text{where} \quad \zeta = \left(\dfrac{\alpha^* {M}^{2}\left(\gamma-1\right)-\beta^* \left(1-{M}^{2}\right)}{1-\gamma{M}^{2}}\right).
\end{aligned}
\end{equation}

The above equations represent the general form of linearised Euler equations in a varying area duct with a non-isentropic mean flow, without neglecting any effects of the entropy perturbations on the acoustic field. By defining wave vector as $\mathbf{W}\left(x^{*}\right) = {\left[ {p}^{+}\left(x^{*}\right) \quad {p}^{-}\left(x^{*}\right) \quad \hat{\epsilon}\left(x^{*}\right) \right]}^{T}$ the equations Eqs.~\eqref{eq:dpplusminus}, \eqref{eq:dsigma} can be represented in matrix notation as,
\begin{equation}\label{eq:DiffMatrixNotation}
   \dfrac{\textrm{d}}{\textrm{d} x^* } \left[\mathbf{W}\left(x^{*}\right)\right] =  \mathbf{C}\left(x^{*}\right) \mathbf{W}\left(x^{*}\right); \quad \textrm{with} \quad \mathbf{C}\left(x^{*}\right) = \begin{bmatrix}
{C}_{11}&{C}_{12}&{C}_{13}\\
{C}_{21}&{C}_{22}&{C}_{23}\\
{C}_{31}&{C}_{32}&{C}_{33}\\
\end{bmatrix}
\end{equation}
The elements of the coefficient matrix, $\mathbf{C}\left(x^{*}\right)$ have the expressions denoted in Eqs.~\eqref{eq:dpplusminus},~\eqref{eq:dsigma}.

It can be observed from Eq.~\eqref{eq:dpplusminus} that the coefficient of $\hat{\epsilon}$ is proportional to the non-dimensional gradients of area and mean temperature ($\alpha^*$, $\beta^*$) and is of the order O$(M^2)$. This analysis assumes low flow Mach numbers - a reasonable assumption for combustors which typically require $M\approx0.2$, to keep the flame attached. Hence the effect of the entropy fluctuations on the acoustic field is neglected, thus decoupling Eq.~\eqref{eq:dpplusminus} from Eq.~\eqref{eq:dsigma}. This simplifies the analysis by retaining only the acoustic waves, such that the coefficient matrix $\mathbf{C}\left(x^{*}\right)$ is reduced from a $3\times3$ matrix to a $2\times2$ matrix and the wave vector $\mathbf{W}$ now only has ${p}^{\pm}$ as unknowns. Thus, Eq.~\eqref{eq:DiffMatrixNotation} reduces to,
\begin{equation}\label{eq:DiffMatrixNotationPureacoustic}
    \dfrac{\textrm{d}}{\textrm{d} x^* } \left[\mathbf{P}\left(x^{*}\right)\right] =  \mathbf{C}\left(x^{*}\right) \mathbf{P}\left(x^{*}\right); \quad \textrm{with} \quad \mathbf{C}\left(x^{*}\right) = \begin{bmatrix}
{C}_{11}&{C}_{12}\\
{C}_{21}&{C}_{22}\\
\end{bmatrix} \quad \textrm{and} \quad \mathbf{\textbf{P}}\left(x^{*}\right) = {\left[ {p}^{+}\left(x^{*}\right) \quad {p}^{-}\left(x^{*}\right) \right]}^{T}.
\end{equation}
Eq.~\eqref{eq:DiffMatrixNotationPureacoustic} represents a system of first-order differential equations which can be numerically solved - a fourth-order Runge-Kutta method is used for this analysis. The numerical implementation of the LEEs is discussed in~\ref{sec:appendixb}. 

Upon integration, a relation between the wave components at the upstream ($x^*_{u}$) and downstream ($x^*_{d}$) ends of the duct (shown in Fig.~\eqref{fig:Figure1}), is obtained and expressed in terms of a transfer matrix $\mathbf{T}$, as follows,
\begin{equation}\label{eq:TransferMatrixFormula}
   \mathbf{P}\left(x^{*}_d\right) =  \left[\mathbf{T}\right]_{2\times2}  \mathbf{P}\left(x^{*}_u\right) \quad \text{where} \quad \mathbf{T} = \begin{bmatrix}
{T}_{11}&{T}_{12}\\
{T}_{21}&{T}_{22}\\
\end{bmatrix}.
\end{equation}
The elements of the transfer matrix can be obtained by applying two sets of independent acoustic boundary conditions, one at either end of the duct (shown in \ref{sec:appendixb}).

The change in acoustic energy flux across a given length of duct with an arbitrary subsonic mean flow can be characterised using the acoustic absorption coefficient ($\Delta$). The rigorous expression for $\Delta$ was shown to be \citep{Gaudron_ICSV_2019, Gaudron_GT&P_2021},
\begin{align}\label{eq:Deltafandg}
\Delta =  1 -  \dfrac{{{{\left(1 + M_d\right)}^2} {\left|{p}^+_d\right|}^{2}}  + \theta \: \Xi \: \mathcal{\xi}^{-1} {{\left(1 - M_u\right)}^2} {\left|{p}^-_u\right|}^{2} }{\theta \: \Xi \:\mathcal{\xi}^{-1} {{\left(1 + M_u\right)}^2}{\left|{p}^+_u\right|}^{2} + {{{\left(1 - M_d\right)}^2}  {\left|{p}^-_d\right|}^{2} }}
\end{align}
where,
\begin{equation*}
                 \refstepcounter{equation}\latexlabel{eq:theta}
                  \refstepcounter{equation}\latexlabel{eq:Xi}
                  \refstepcounter{equation}\latexlabel{eq:xi}
\theta = \dfrac{{A}_{u}}{{A}_{d}}, \quad                  
\Xi = \sqrt {{\overline{{T}}_{u}}/{\overline{{T}}_{d}}}, \quad \text{ and }\quad \mathcal{\xi} = {\overline{p}_{u}}/{\overline{p}_{d}}. \tag{\ref*{eq:theta} - \ref*{eq:xi}}
\end{equation*}

For a duct with given area and mean temperature profiles, the mean flow can be solved to determine the Mach number distribution and the mean pressure ratio, $\xi$. With the mean flow solution established, the first-order system of differential equations given by Eq.~\eqref{eq:dpplusminus} can then be solved for given acoustic boundary conditions to fully determine the elements of transfer matrix, given in Eq.~\eqref{eq:TransferMatrixFormula}. The acoustic wave amplitudes can be obtained from Eq.~\eqref{eq:DiffMatrixNotationPureacoustic}, and used in Eq.~\eqref{eq:Deltafandg} to compute $\Delta$. The acoustic absorption coefficient ($\Delta$), can also be predicted from the WKB method based solution \citep{Yeddula_JSV_2020} for the acoustic field. The general semi-analytical solution derived in this work is applicable to any arbitrary area and mean temperature variation, and is described in Section ~\ref{sec:Analytical}.
\section{Analytical model for the acoustic absorption coefficient}\label{sec:Analytical}
For varying area ducts sustaining a mean flow which may be non-isentropic, a semi-analytical solution based on an iterative WKB method was presented in our previous work \citep{Yeddula_JSV_2020}. It requires the frequency to be ``large'' in one sense, while also being low enough for only plane waves to propagate in the duct. The final solution for the acoustic pressure and velocity is expressed as,
\begin{equation}\label{eq:phatuhatfinalIsentropic}
\hat{p}\left(x^*,\omega\right) = {{C}}^{+}{{\mathcal{P}}}^{+}\left(x^*,\omega\right) + {{C}}^{-}{{\mathcal{P}}}^{-}\left(x^*,\omega\right), \quad \text{and} \quad     \hat{u}\left(x^*,\omega\right) = \dfrac{1}{\overline{\rho}\overline{c}}\left({\mathcal{B}}^{+}{C}^{+}{\mathcal{P}}^{+} - {\mathcal{B}}^{-}{C}^{-}{\mathcal{P}}^{-}\right),
\end{equation}
with $C^{\pm}$ constants determined from boundary conditions,
\begin{equation}\label{eq:FinalAnalyticalSolutionNonIsentropic}
    {{\mathcal{P}}}^{\pm} = {\left(\dfrac{{A}_{u}}{A}\right)}^{{1}/{2}}{\left(\dfrac{{\overline{T}}_{u}}{\overline{T}}\right)}^{{1}/{4}}\left[\dfrac{\text{exp} \left(\dfrac{{M}^{2}}{2} \mp {M} + \dfrac{\gamma}{2}\displaystyle{\int _{{x_u^*}}^{x^*}{(\alpha^* {M}^{2} \mp \beta^* M)\text{d}\breve{x}}}\right)}{\text{exp}\left(\dfrac{{M}_{u}^{2}}{2} \mp {M}_{u}\right)}\right] \text{exp}\left(\int _{ {x_u^*} }^{x^*}{ \text{i} \textrm{He}\left( \dfrac{\mp 1}{1 \pm M} \pm \dfrac{\Phi_{NI}^{\pm} }{2 \text{He}^2} \right) \text{d}\breve{x}} \right),
\end{equation}
\begin{equation}\label{eq:B1plusminusNonIsentropic}
    {\mathcal{B}}^{\pm} = \dfrac{\text{i} \text{He}\left(1-\gamma M^2\right) \pm \dfrac{\beta^*}{4} \pm \dfrac{\alpha^*}{2} -\alpha^* M + \dfrac{\beta^* M}{2}\left(1+\gamma\right) \pm \dfrac{\beta^* M^2}{4}\left(3\gamma - 7\right) \pm \dfrac{\alpha^* M^2}{2}\left(3-2\gamma\right)}{\text{i} \text{He}\left(1-\gamma M^2\right) + \left(\beta^* - 2\alpha^* \right) M},
\end{equation}
where $\breve{x}$ is a dummy variable. The expression for $\Phi_{NI}^{\pm}$ in Eq.~\eqref{eq:FinalAnalyticalSolutionNonIsentropic} is given by,
\begin{equation}
 \begin{aligned}\label{eq:PhiIsentropic}
  {\Phi_{NI}^{\pm}} & =  
   \dfrac{{\alpha^*}^2}{4} + \dfrac{3{\beta^*}^2}{16} + \dfrac{{\alpha^*} {\beta^*}}{2} \mp M \left({\alpha^*}^2 + \dfrac{{\beta^*}^2}{2}\left(1 - \gamma \right) - \dfrac{{\alpha^*} {\beta^*}}{2} \left(3-\gamma\right) \right) + \dfrac{\text{d}{\alpha^*}}{dx}\left( \dfrac{1}{2} \pm M\right) + \dfrac{1}{2}\dfrac{\text{d}{\beta^*}}{dx}\left(\dfrac{1}{2} \pm M\left(\gamma -1 \right) \right).
 \end{aligned}
 \end{equation}
 
 The above solution is considered semi-analytical as Eq.~\eqref{eq:FinalAnalyticalSolutionNonIsentropic} requires the integrals to be numerically evaluated. This WKB method based solution also inherently assumes the frequency to be larger than a minimum value, with the lower frequency limit presented in \citep{Yeddula_JSV_2020} requiring $\dfrac{\Phi_{NI}^{\pm}}{\text{He}^2} \ll 1$.
 
 On comparing Eqs.~\eqref{eq:phatuhatDefinitionDokumaci},~\eqref{eq:phatuhatfinalIsentropic}, and defining the upstream reflection coefficient as $R_u = {p_u^+}/{p_u^-}$, the constants $C^{\pm}$ are obtained in terms of $R_u$ as,
\begin{equation*}
             \refstepcounter{equation}\latexlabel{eq:CPlusMinus}
                  \refstepcounter{equation}\latexlabel{eq:OutletToInlet}
    \dfrac{C^{\pm}}{p^{-}_{u}} = \dfrac{\left(\mathcal{B}^{\mp}_{u}\pm1\right)R_u + \left(\mathcal{B}^{\mp}_{u}\mp1\right)}{\left(\mathcal{B}^{+}_{u} + \mathcal{B}^{-}_{u}\right)}, \quad \textrm{and} \quad     {{p}^{\pm}_d} = \dfrac{1}{2}\left[{C^+} \: {\mathcal{P}_d^+} \left(1 \pm \mathcal{B}^+_d\right) + {C^-} \: {\mathcal{P}_d^-} \left(1 \mp \mathcal{B}^-_d\right)\right],
                 \tag{\ref*{eq:CPlusMinus}, \ref*{eq:OutletToInlet}}
\end{equation*}
since using Eq.~\eqref{eq:FinalAnalyticalSolutionNonIsentropic} gives $\mathcal{P}_{u}^{\pm} = 1$ at the upstream end of the duct. The acoustic absorption coefficient $\Delta$, given by Eq.~\eqref{eq:Deltafandg}, is therefore estimated using the analytical expressions for ${p}^{\pm}_{d}$, given by Eq.~\eqref{eq:OutletToInlet}.
\section{Setting the validation test case}\label{sec:Setting the validation test case}
The analysis described in Section \ref{sec:Analytical} is applicable to ducts with any area and mean temperature profiles, provided that $\dfrac{\Phi_{NI}^{\pm}}{\text{He}^2}\ll1$. In this section, a conical duct geometry with a linear stream-wise mean temperature gradient, shown in Fig.~\ref{fig:Figure1}, is considered. The radius of the duct is denoted by $r$, from which it follows that,
\begin{align}\label{eq:ConicalAreaProfile}
    A\left(x^*\right) = \: \pi r_u^2{( 1 + a^*\:x^*)}^{2}, \quad \overline{T}\left(x^*\right)\:= \overline{T}_{u}{\left(1 + b^* x^*\right)}; \quad \text{where} \quad a^*\;= \; \dfrac{r_d - r_u}{r_u}, \quad b^*\;= \; \dfrac{\overline{T}_d - \overline{T}_u}{\overline{T}_u}. 
\end{align}
The non-dimensional area and mean temperature gradient parameters, $\alpha^*$, $\beta^*$, and their corresponding spatial derivatives become,
        \begin{equation*}
                  \refstepcounter{equation}\latexlabel{eq:AlphaGradientRelation}
                  \refstepcounter{equation}\latexlabel{eq:DalphaRelation}
                  \refstepcounter{equation}\latexlabel{eq:BeetaGradientrelation}
                  \refstepcounter{equation}\latexlabel{eq:Dbetarelation}                  
\alpha^* = \dfrac{1}{A}\dfrac{\text{d}A}{\text{d}x^*} = \dfrac{2 a^*}{(1+a^*\:x^*)} \; \Rightarrow \; \dfrac{\text{d}\alpha^*}{\text{d}x^*} = \;-\dfrac{{\alpha^*}^{2}}{2}; \quad \beta^* = \dfrac{1}{\overline{T}}\dfrac{\text{d}\overline{T}}{\text{d}x^*} = \dfrac{ b^*}{(1+b^*\:x^*)} \; \Rightarrow \; \dfrac{\text{d}\beta^*}{\text{d}x^*} = \;-{\beta^*}^{2}.
                 \tag{\ref*{eq:AlphaGradientRelation} - \ref*{eq:Dbetarelation}}
       \end{equation*}
For this particular case of a conical duct with linear stream-wise mean temperature variation, a simplified analytical solution was presented in \cite{Yeddula_JSV_2020} to determine the acoustic response by evaluating the integrals in Eq.~\eqref{eq:FinalAnalyticalSolutionNonIsentropic}. This brings down the computational effort and makes the solution at any location depend only on the local mean flow properties and frequency. The main results are summarised in \ref{sec:appendixc}, with the final expression for ${\mathcal{P}}^{\pm}_d$, given by Eq.~\eqref{eq:P1plusminusNonIsentropicConical_Appendix} showing that it is only a function of mean flow properties at $x^*_u,\: x^*_d$ and $\textrm{He}_u$ (He$_u = \omega L / \overline{c}_u$). These expressions can be used to estimate $p^{\pm}_{d}$ analytically using Eq.~\eqref{eq:OutletToInlet}. The corresponding lower frequency limit, given by Eq.~\eqref{eq:PhiIsentropic}, can be recast in terms of a critical Helmholtz number at the upstream end (He$_\textrm{cr,\:u}$),       
\begin{equation}\label{eq:FrequencyLimitLower}
    \textrm{He}_u \gg \textrm{He}_{\textrm{cr},\:u} = \textrm{max} \left[\dfrac{L \overline{c}}{\overline{c}_u}\sqrt{\left| -\dfrac{{b^*}^2}{16} + 2a^* b^*\right|}, \:\dfrac{L \overline{c}}{\overline{c}_u} \sqrt{\left| \dfrac{-1}{16}\dfrac{{b^*}^2}{(1+b^*)^2} + \dfrac{2a^* b^*}{(1+a^*)(1+b^*)}\right|}\right].
\end{equation}

The WKB solution used for estimating $\Delta$ is valid for ducts sustaining planar acoustic waves, which means that higher order acoustic modes must be cut-off. This sets a restriction on the upper limit on the frequency that can be considered. For the case of ducts with circular cross-sectional area, the first non-planar modes occur for frequencies higher than He$_{upper,\:u}$,
\begin{equation}
    \textrm{He}_{upper,\:u} = \textrm{{min}}\left[\frac{1.841 \:L\: \overline{c}}{r \:\overline{c}_u}\right] \label{eq:FrequencyLimitUpper}.
\end{equation}
The ``min'' and the ``max" functions in Eqs.~\eqref{eq:FrequencyLimitLower},~\eqref{eq:FrequencyLimitUpper} gives the minimum and maximum values for the limits of the frequency inside the duct.
\begin{figure}[H]  
\small
\def\svgwidth{1.2\textwidth}
\begin{minipage}{.45\linewidth}
\raggedleft
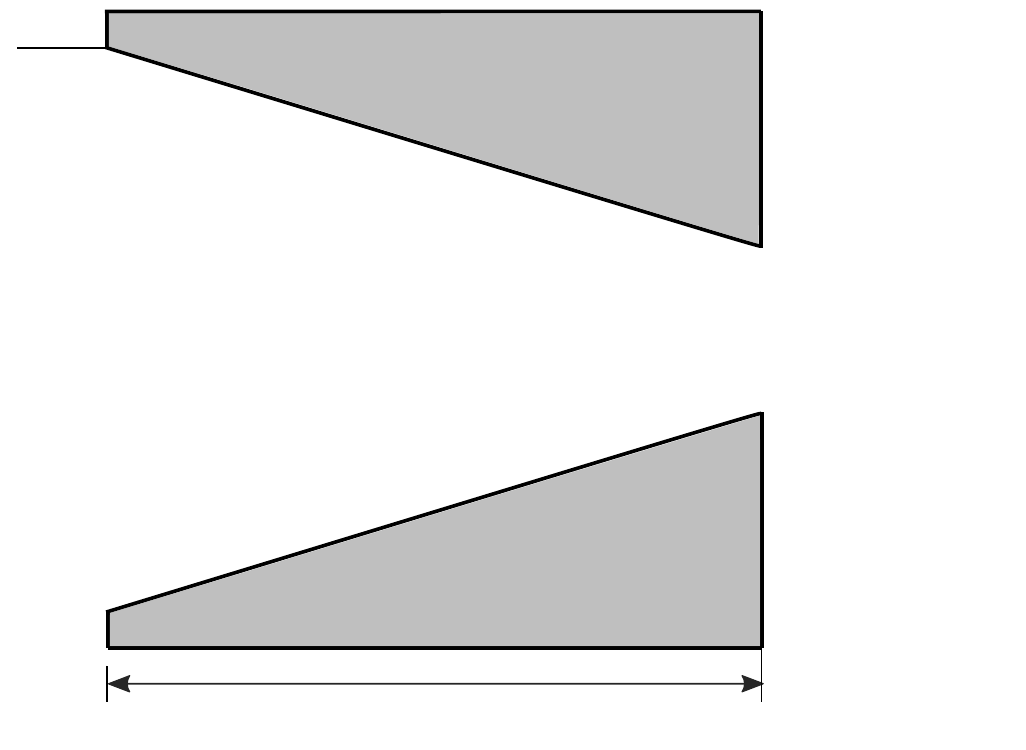
\end{minipage}
\hfill
\begin{minipage}{.5\linewidth}
\raggedright
        \begin{tabular}{c l l}
        \toprule
     \textbf{Mean flow} & Upstream & Downstream\\
      \cmidrule(lr){2-2}\cmidrule(lr){3-3}
      $r$ & 240 \textrm{mm} & 168 \textrm{mm}\\
    \addlinespace
        $ \left|\alpha^*\right|$ & 1.2 & 1.71\\
      \addlinespace
      $M = \dfrac{\overline{u}}{\overline{c}}$ & 0.15& 0.273 \\
      \addlinespace
        $ \overline{T}$ & 1600K & 1120K\\
      \addlinespace
        $ \left|\beta^*\right|$ & 0.5 & 0.67\\
      \addlinespace    
      \hdashline
            \textbf{Frequency limits} & $\textrm{He}_u$ & $f_u$\\
      \cmidrule(lr){2-2}\cmidrule(lr){3-3}
      $\textrm{Lower limit}$ Eq.~\eqref{eq:FrequencyLimitLower} & 0.26 & 67\textrm{Hz}\\
      \addlinespace
            $ \textrm{Upper limit}$ Eq.~\eqref{eq:FrequencyLimitUpper}  &6.8& 1735\textrm{Hz}\\
      \addlinespace
      \bottomrule
        \end{tabular}
        \centering
        \captionsetup{width=.8\linewidth}
        \captionsetup{justification=centering}
    \end{minipage}%
    \centering
        \caption{Schematic of the $L=0.5$m long conical-shaped  duct ($a^* = -0.3$) sustaining a non-isentropic mean flow ($b^* = -0.3$). Other mean flow parameters and calculated frequency limits are also presented. Frequency limits are specified in terms of both Helmholtz numbers and frequencies $\left(f = \dfrac{\textrm{He} \:\overline{c}_u }{2 \pi L} \right)$.}
        \label{fig:Fig2}
\end{figure}
A conical duct of length $0.5$m, with radii of $240$mm and $168$mm at the upstream and downstream ends is considered. The duct sustains a subsonic mean flow in the presence of a linear temperature gradient. From the upstream to the downstream end the temperature varies from 1600K to 1120K, while the flow Mach number varies from $0.15$ to $0.273$. The schematic in Fig.~\ref{fig:Fig2} shows the duct geometry, mean flow parameters along with the corresponding frequency limits given by Eqs.~\eqref{eq:FrequencyLimitLower},~\eqref{eq:FrequencyLimitUpper}. It can be observed that there exists a range of frequencies (67Hz to 1735Hz) in which the model can be applied for acoustic field estimation without significant deviation from the numerical results \citep{Yeddula_JSV_2020, Li_JSV_2017}.

 Finally, the predicted values of $p^{\pm}_{d}$ are coupled with Eq.$~\eqref{eq:Deltafandg}$ to estimate the acoustic absorption coefficient ($\Delta$) in conical ducts with linear stream-wise mean temperature gradient. The model estimates for $\Delta$ are compared to the calculations of $\Delta$ using the numerical LEEs, with the comparisons now presented in Section \ref{sec:Results}. 
\section{Results}\label{sec:Results}
\subsection{Effect of frequency (He) and upstream reflection coefficient (R$_\textrm{u}$): }
\begin{figure}[t]
\centering
\includegraphics[clip, trim=1.2cm 1cm 0.6cm 0cm, width=\textwidth, height = 4.7in]{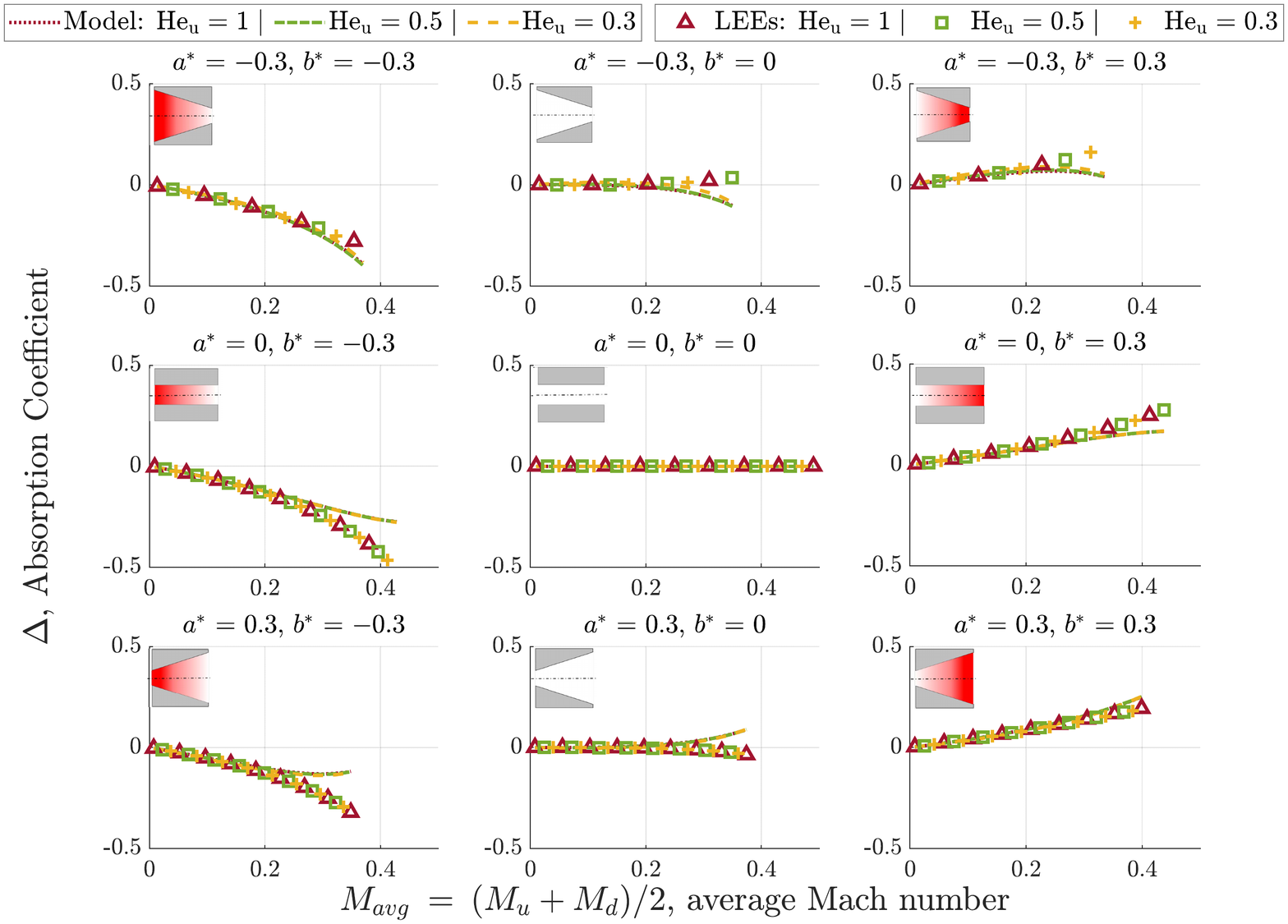}
\caption{Comparison of $\Delta$ obtained using analytical model (represented by lines) and numerical LEEs (represented by markers) as a function of average Mach number ($M_{u}$ + $M_{d}$)/2, for a conical duct with linear mean temperature gradient and an anechoic boundary at the upstream end ($|R_u|\:=\:0$). Top row: $a^* = -0.3$, centre row: $a^* = 0$, bottom row: $a^* = 0.3$; left column: $b^* = -0.3$, middle column: $b^* = 0$, right column: $b^* = 0.3$.}
\label{fig:Figure2}
\end{figure}

The acoustic absorption coefficient across the conical duct with linear mean temperature variation is now compared as calculated using the numerical simulations and the analytical model, for a range of average flow Mach numbers and for three different values of $a^*$ and $b^*$. The results, shown in Fig.~\ref{fig:Figure2}, are presented for three different values of frequency (in terms of Helmholtz number at the upstream end, He$_{u}$), for an upstream anechoic boundary condition ($|R_u|=0$). 

In Fig.~\ref{fig:Figure2}, as the sign of area coefficient $a^*$, varies from negative to positive (top to bottom), the duct changes from converging to a diverging duct. Similarly, the mean temperature coefficient $b^*$ changes sign from left to right, with negative (respectively positive) $b^*$ representing a decreasing (respectively increasing) mean temperature profile, with heat loss to (respectively gain from) the surroundings. Thus, negative value of $b^*$ correspond to cooled ducts, while a positive value of $b^*$ correspond to heated ducts. These variations in $a^*$ and $b^*$ are illustrated using a graphical representation of the duct profile and mean temperature distribution over each plot in Fig.~\ref{fig:Figure2}. A flow Mach number of up-to a maximum of $0.5$ is considered to examine the model behaviour at increased subsonic flow Mach numbers. 

From Fig.~\ref{fig:Figure2}, it can be observed that, for an upstream anechoic boundary, both the numerical and the model results are independent of the Helmholtz number in the frequency range of interest. Furthermore, when $M_{avg}$ is less than $0.25$, the model estimates of the acoustic absorption coefficient, $\Delta$, are in close agreement with the validation simulations.
 
\begin{figure}[t]
\centering
\includegraphics[clip, trim=1.2cm 1cm 0.6cm 0cm, width=\textwidth, height = 4.7in]{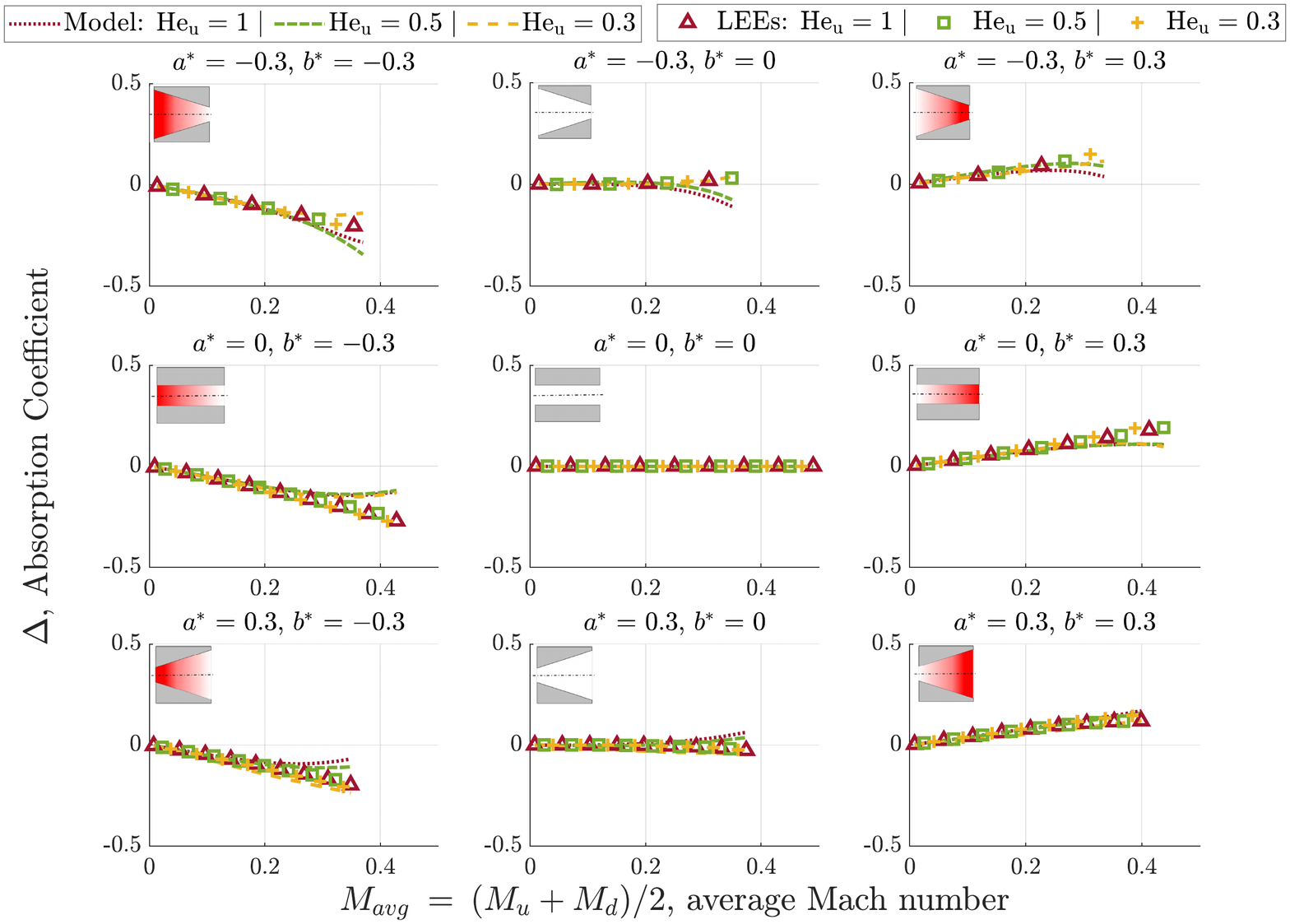}
\caption{Comparison of $\Delta$ obtained using the model (represented by lines) and numerical LEEs (represented by markers) as a function of average flow Mach number ($M_{u}$ + $M_{d}$)/2, for a conical duct with linear mean temperature gradient and non-anechoic boundary at the upstream end with $|R_u|\:=\:0.5$. Top row: $a^* = -0.3$, centre row: $a^* = 0$, bottom row: $a^* = 0.3$; left column: $b^* = -0.3$, middle column: $b^* = 0$, right column: $b^* = 0.3$.}
\label{fig:Figure3}
\end{figure} 

Fig.~\ref{fig:Figure3} compares the model and numerical LEEs calculations of $\Delta$ for a non-anechoic inlet boundary condition with $|R_u|=0.5$. It should be noted that Eq.~\eqref{eq:Deltafandg} contains only the absolute value of the upstream reflection coefficient, and thus the phase ($\angle{R_u}$) has no effect. For $M_{avg}$ less than $0.25$, the model predictions again agree well with the validation simulations. However, the numerical estimates remain independent of frequency, while the model predictions start to deviate at He$_u=0.3$. This behaviour can be attributed to the violation of lower frequency limit which requires He$_u$ $>$ He$_\textrm{cr} = 0.43$. 

Figs.~\ref{fig:Figure2} and~\ref{fig:Figure3} suggest that the value of $\Delta$ is independent of frequency, provided the lower frequency limit criterion, set by Eq.~\eqref{eq:FrequencyLimitLower}, is satisfied. For the mean flow parameters considered, the absolute value of $\Delta$ for the anechoic case in Fig.~\ref{fig:Figure2} is larger than its non-anechoic counterpart in Fig.~\ref{fig:Figure3}, i.~e., $|\Delta_{R_u\:=\:0}|\;> |\Delta_{R_u\:=\:0.5}|$. This was checked to hold for a majority of duct area and mean temperature profiles, and an example result is presented in \ref{sec:appendixd}. Further, a simplified mathematical analysis also showed that the acoustic absorption coefficient ($\Delta$) is independent of frequency (He$_u$), and its absolute value ($|\Delta|$) maximises for anechoic upstream boundary condition $|R_u| = 0$. This simplified analysis assumes low flow Mach numbers (retaining only terms up to order O$(M)$) and requires the frequency criteria set by Eqs.~\eqref{eq:FrequencyLimitLower},~\eqref{eq:FrequencyLimitUpper} to be satisfied. The final results are presented in \ref{sec:appendixF}.

These results suggest that to identify extrema of $\Delta$, which correspond to regions of maximum acoustic absorption or generation, it is reasonable to study the case of anechoic boundary at the upstream end for a particular frequency, say He$_u=1$. Thus, the dependence of $\Delta$ on acoustic parameters (He, $R_u$) is eliminated, and the study can be confined to analysing the effects of mean flow parameters on $\Delta$.
\begin{figure}[t]
\centering
\includegraphics[angle=0, clip, trim=1.5cm 1cm 0.6cm 0cm, width=1\textwidth, height = 4.4in]{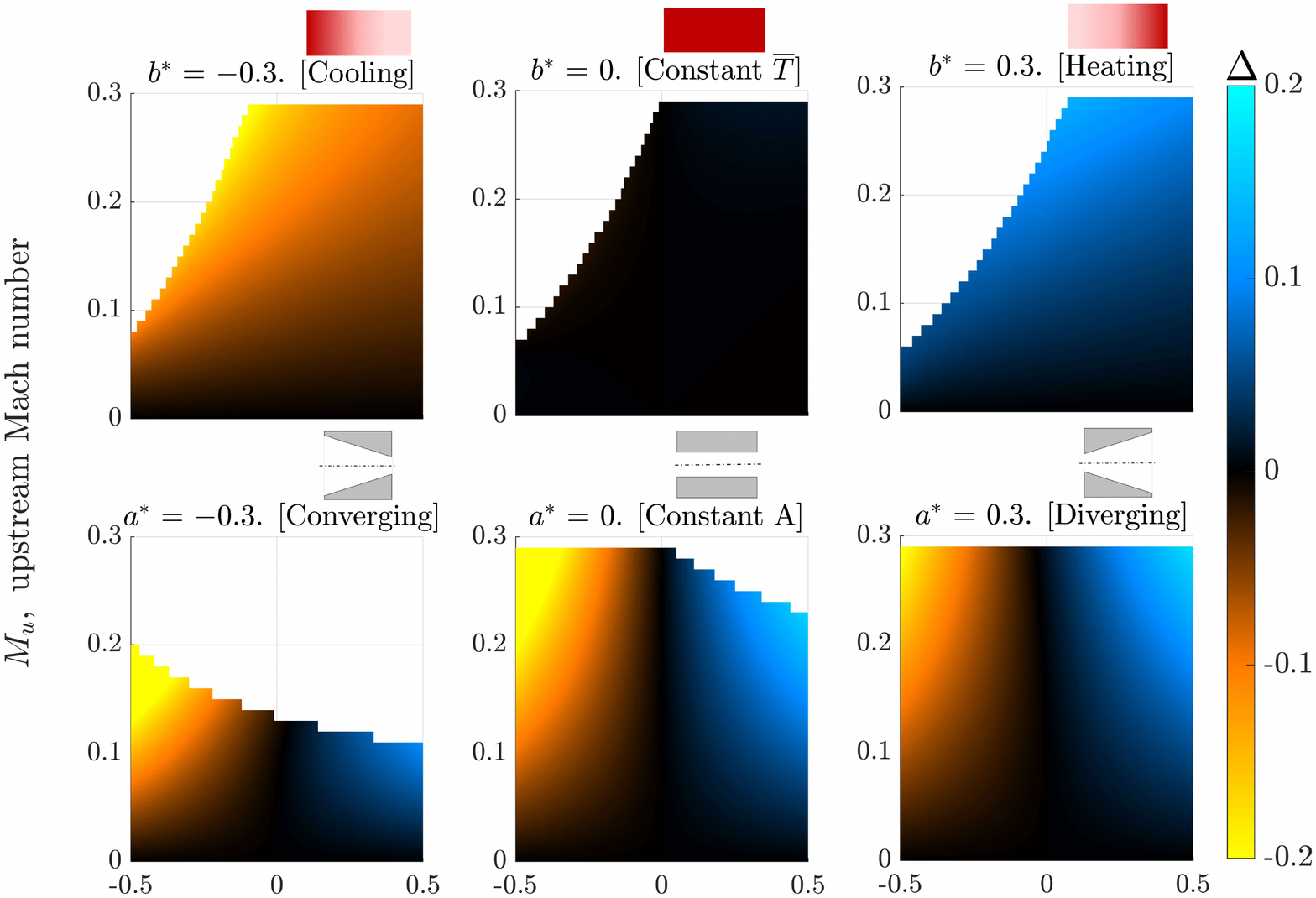}
\caption{Contours of $\Delta$ obtained by the model (for He$_u$ = 1) as a function of the flow Mach number at the inlet $M_{u}$, for a conical duct with anechoic boundary at the upstream end ($|R_u|\:=\:0$) and a linear mean temperature gradient. The top row plots are for a given $b^*$ (top left: $b^* = -0.3$, top centre: $b^* = 0$, top right: $b^* = 0.3$) with varying $a^*$, while the bottom row plots are for a given $a^*$ (bottom left: $a^* = -0.3$, bottom centre: $a^* = 0$, bottom right: $a^* = 0.3$) with varying $b^*$.}
\label{fig:Figure4}
\end{figure} 

\subsection{Effect of mean flow parameters}
The effect of mean flow parameters - area variation ($a^*$), mean temperature variation ($b^*$) and upstream flow Mach number ($M_u$) - on $\Delta$ is shown in Fig.~\ref{fig:Figure4}, which plots the model's estimate of $\Delta$ as a function of $M_u$, for a given $b^*$ (respectively $a^*$) on the top row (respectively bottom row) and varying $a^*$ (respectively $b^*$) from -0.5 to 0.5. The top row of Fig.~\ref{fig:Figure4} reveals an important finding. For a negative stream-wise mean temperature gradient, the absorption coefficient is always negative, corresponding to acoustic energy generation across the duct. Conversely, for a positive stream-wise mean temperature gradient, the absorption coefficient is always positive, corresponding to acoustic energy absorption across the duct. When there is a constant mean temperature maintained in the duct, for any value of area variation, $\Delta$ is close to zero, i.e., there is negligible generation or absorption of acoustic energy. This behaviour is further analysed in Section \ref{sec:EnergyAnalysis}. At low flow Mach numbers, the simplified mathematical analysis presented in \ref{sec:appendixF} also confirms that the generation and absorption of acoustic energy occurs in cooled and heated ducts, respectively, with negligible dependence on area variation - for any arbitrary area profile considered.

Regions with high acoustic energy absorption or generation can also be observed in Fig.~\ref{fig:Figure4}, and are mostly confined to high flow Mach number regimes. In general, the non-dimensional mean temperature variation, $b^*$, has a stronger effect on $\Delta$ than the non-dimensional area variation, $a^*$. Fig.~\ref{fig:Figure4} suggests that, for a given $b^*$, there exists a region of $a^*$ values which maximise $\left|\Delta\right|$. An optimisation search for the global extrema was therefore performed, based on a genetic algorithm, with the results discussed in \ref{sec:appendixE}. The main findings are that certain mean flow parameters result in either a significant acoustic energy damping ($\sim 60\%$) or a significant acoustic energy generation ($\sim 110\%$).

Practical combustors often encounter a decreasing stream-wise mean temperature, due to heat loss and the input of cold dilution air. For a given combustor geometry, the area profile is fixed, and therefore by adjusting the mean temperature profile, the acoustic energy generated can be controlled. For example, for $a^*=-0.3$, by reducing $b^*$ by 20\% from -0.5 to -0.4, the acoustic energy generated can be reduced by $23\%$ ($\Delta$ from -0.245 to -0.19). This inverse linear dependence of $\Delta$ on stream-wise mean temperature variation, is further investigated using energy analysis discussed in Section~\ref{sec:EnergyAnalysis}.
\section{Energy Analysis}\label{sec:EnergyAnalysis}
To gain further insights into the findings regarding the stream-wise mean temperature gradient effect, as discussed in Sec~\ref{sec:Results}, an acoustic energy analysis is now performed. Using the definitions of Morfey \citep{Morfey_JSV_1971}, the acoustic energy flux ${\textbf{N}^*}$, is given in terms of total enthalpy fluctuations ($J^*$) and mass fluctuations ($m^*$) for the current 1D case as,
\begin{equation}
\begin{aligned}
   & \; \textbf{N}^* = {J^* m^*}; \quad \textrm{with} \; J^* = \dfrac{p'}{\overline{\rho}} + \overline{u} {u'} \quad \textrm{and} \quad m^* = \overline{\rho}  u' + \dfrac{M}{\overline{c}}{p'}.
\end{aligned}
\end{equation}
By multiplying Eq.~\eqref{eq:NormalisedMassEnergy} with $\overline{\rho} \: J^*$, Eq.~\eqref{eq:NormalisedMomentum} with $\overline{u} \:m^*$, adding the two equations and rearranging, we obtain
\begin{equation}\label{eq:EnergyCorollary}
\begin{aligned}
    &\dfrac{ \partial }{ \partial t }\overbrace{\left[\dfrac{{p'}^2}{2\overline{\rho} \:\overline{c}^2} + {p'} {u'}\dfrac{M}{\overline{c}} + \dfrac{1}{2}\overline{\rho} {{u'}}^2\right]}^{\textbf{E}^*} + 
                \dfrac{ \partial }{ \partial x }\overbrace{\left[{ {{p'}}^2 \dfrac{M}{\overline{\rho}\: \overline{c}}  + {p'} {u'}\left(1 + M^2\right) + \overline{\rho}\: \overline{u}{{u'}}^2 }\right]} ^{\textbf{N}^*}\\ & 
                =\dfrac{1}{1 - \gamma M^2} \Bigg[ \Bigg. \dfrac{-{{p'}}^2 \overline{u}}{\gamma \overline{p}} \left(\beta^* \left(\gamma-1\right) + \alpha^* \right) - {p'} {u'} \alpha^* +  \dfrac{{{p'}} M^2 \overline{u}}{\gamma \overline{p}} \left(\beta^* \left(1-2\gamma\right) + \alpha^* \left(\gamma^2 + \gamma-1\right)\right) \\ & + {p'} {u'} M^2 \left( 3\alpha^* \left(\gamma - 1\right) + \beta^* \left(\dfrac{5}{2}-3\gamma + \gamma \dfrac{\alpha^* + \beta^*}{2} M^2\right)\right) - \overline{\rho}\: \overline{u}{ {{u'}}^2} \left(\dfrac{\beta^*}{2}\left(1-M^2\right)\left(2-\gamma\right)  + \alpha^* \left(1-M^2 \left(2\gamma-1\right)\right) \right)\Bigg. \Bigg].
\end{aligned}
\end{equation}
Note that the contribution of entropy fluctuations to the disturbance energy is neglected, consistent with the low Mach number flow assumption. Eq.~\eqref{eq:EnergyCorollary} equation can be expressed as,
\begin{equation}\label{eq:EnergyMorfey}
        \dfrac{ \partial \textbf{E}^* }{ \partial t } + 
                \dfrac{ \partial \textbf{N}^* }{ \partial x }
                = \textbf{G},
\end{equation}
where \textbf{G} is given by the right hand side of Eq.~\eqref{eq:EnergyCorollary}.

Eq.~\eqref{eq:EnergyMorfey} gives the acoustic energy balance of the duct system under consideration; $\textbf{E}^*$ signifies the acoustic energy density while $\textbf{G}$ corresponds to the acoustic energy generation rate. Thus, $\textbf{G} > 0$ signifies acoustic energy generation and \textbf{G}$<0$ corresponds to acoustic energy absorption.

In the limit of very low Mach number flow in ducts without area variation ($\alpha^* = 0$) the expression for \textbf{G} takes the form,
\begin{equation}\label{eq:GenerationTerm}
\textbf{G} \sim  -\beta^* \overline{u}\: \left[\dfrac{{p'}^2 }{\gamma \overline{p}} \: \left(\gamma - 1 \right) + \overline{\rho}  {u'}^{2}\right].  
\end{equation}
From this expression, it can be seen that acoustic energy absorption (respectively generation) always occurs for ducts with positive (respectively negative) mean temperature gradients, as the term inside the square brackets in Eq.~\eqref{eq:GenerationTerm} is always positive. 

As most practical duct flow systems exhibit a decreasing mean temperature profile due to heat loss to the surroundings, Eq.~\eqref{eq:GenerationTerm} suggest a generation of acoustic energy. This explains the nature of results observed in Section \ref{sec:Results} and also in \ref{sec:appendixF}, and may be of special relevance to thermoacoustic systems with embedded heat exchangers.
\section{Conclusions}\label{sec:Conclusions}
This paper presents an analytical framework for estimating the acoustic absorption coefficient, $\Delta$, in a smoothly varying area duct sustaining a mean flow with a stream-wise varying mean temperature. This exposed the direct linkage between heated/cooled ducts and acoustic energy damping/generation. This is achieved by firstly establishing a numerical mechanism to solve the linearised Euler equations for the acoustic wave amplitudes propagating in the downstream and upstream directions, with a low Mach number flow assumption, allowing the effect of the entropy fluctuations on the acoustic field to be neglected. The WKB method based solution presented in our previous work \citep{Yeddula_JSV_2020} is used to predict the acoustic response of the duct. The acoustic absorption coefficient, $\Delta$, is evaluated using both the numerical and analytical approaches. The analytical predictions were found to closely match the validation simulations, provided the lower frequency limit set by the analytical solution is satisfied and the average of the flow Mach numbers at the upstream and downstream ends is less than $\sim0.25$. The acoustic absorption coefficient was found to depend very little on frequency. It was also observed that the magnitude of $\Delta$ was typically maximised when the upstream boundary was anechoic rather than non-anechoic. $\Delta$ was seen to depend more strongly on the non-dimensional mean temperature variation than on the non-dimensional area variation. Furthermore, for a given area profile, $\Delta$ was observed to be positive, corresponding to acoustic absorption, for heated ducts, and negative, corresponding to acoustic generation, for cooled ducts. This was explained by performing an energy analysis and a simplified mathematical analysis for the case of a uniform area duct sustaining very low Mach number flow. As practical combustors encounter a decreasing temperature across the duct, efforts to reduce acoustic energy generation can be achieved by either altering the mean temperature profile or by choosing the particular value of $a^*$ that minimizes $|\Delta|$.
\section{Acknowledgements}
The authors would like to acknowledge the European Research Council (ERC) Consolidator Grant AFIRMATIVE (2018-2023) and Inlaks Shivdasani foundation for supporting the current research. Aimee Morgans would like to gratefully acknowledge the influence of Shôn Ffowcs Williams (1935-2020) on her career. Shôn supervised her Masters project at Cambridge in his final year before retirement. He was her gateway to the world of academia and research, as well as her first ever co-author. His enthusiasm, sense of humour and support of her early career had a long-lasting impact.

\bibliographystyle{elsarticle-num}
\bibliography{Bibliography.bib}
\appendix
\section{Derivation of mean and perturbed flow conservation equations}\label{sec:appendixa}
Substituting for the flow variables as defined in Eqs.~\eqref{eq:LinearisingFlowVariables}, into the conservation equations given by Eqs.~\eqref{eq:massCons}~-~\eqref{eq:enrCons} results in,
\begin{equation}\label{eq:MassLinearise}
         A \dfrac{ \partial  }{ \partial t }\left(\overline{\rho} + \rho'\right)  + \dfrac{\partial}{\partial x^*}  \left(\overline{\rho} + \rho'\right) A\left(\overline{u} + u'\right) = 0;
        \end{equation}
        \begin{equation}\label{eq:MomentumLinearise}
            \dfrac{\partial}{ \partial t}\left(\overline{u} + u'\right)+\left(\overline{u} + u'\right)\dfrac{\partial}{\partial x^*}\left(\overline{u} + u'\right) + \dfrac{1}{\overline{\rho} + \rho'}\dfrac{\partial }{\partial x^*}\left(\overline{p} + p'\right) = 0;
        \end{equation}        
        \begin{equation}\label{eq:EnergyLinearise}
        \dfrac{ \partial }{ \partial t } {(\overline{s} + s')} +(\overline{u} + u')\dfrac{\partial}{\partial x}\left(\overline{s} + s'\right) = \dfrac{{R}_{g}}{\overline{p} + p'}{\overline{\dot{Q}}}. 
        \end{equation}

Let $\eta$ be a quantity of the order of the ratio of fluctuating component to the mean time-averaged component of any flow variable, for example $\eta \approx p'/\overline{p}$. The linearisation process assumes the fluctuating terms $(\;)'$ to be much smaller than the mean time-averaged quantities $\overline{(\;)}$, that is $\eta \ll 1$. Thus, in writing out the conservation equations for the mean time-averaged flow, the contributions from the time-dependent fluctuating terms can be separated. This gives: 
\begin{equation}\label{eq:TimeAveragedEquations}
\dfrac{\textrm{d}\left(\overline{\rho}A\overline{u}\right)}{\textrm{d} x^*}   = 0, \quad
\overline{\rho}\overline{u}\dfrac{\textrm{d}\overline{u}}{\textrm{d} x^*} + \dfrac{\textrm{d}\overline{p}}{\textrm{d} x^*}  =0, \quad 
\overline{u}\dfrac{\textrm{d} \overline{s}}{\textrm{d} x^*}  = \dfrac{{R}_{g}}{\overline{p}}{\overline{\dot{Q}}}.
\end{equation}
In writing out the perturbation equations only terms of order $\eta$ are considered, and terms of the order O$\left(\eta^2\right)$ are neglected. This results in Eqs.~\eqref{eq:LEE1NIfr}~-~\eqref{eq:LEE3NIfr}, where the mean conservation equations are further used to include the mean time-averaged quantities inside the spatial derivative.

\section{Numerical procedure for solving the linearised Euler equations}\label{sec:appendixb}
The linearised Euler equations given by Eq.~\eqref{eq:dpplusminus} are solved assuming a semi-infinite boundary condition at $x^* = 1$ \citep{Dowling_PCI_2015}. This results in an initial value problem which is solved using a fourth-order Runge-Kutta method. The first-order system of ODEs are as follows,
\begin{equation}\label{eq:SysOfODEs}
\begin{aligned}
 \dfrac{\text{d}{p}^{+}}{\text{d}x^*} & = C_{11} {p}^{+} + C_{12} {p}^{-}, \;\; \& \;\;  \dfrac{\text{d}{p}^{-}}{\text{d}x^*} = C_{21} {p}^{+} + C_{22} {p}^{-}, \quad \textrm{subjected to} \; {p}^{+}_{u} \; \textrm{and} \; {p}^{-}_{u} \; \textrm{at the inlet,}
\end{aligned}
\end{equation}
where $R_u$ is the reflection coefficient at the upstream boundary $R_u = \dfrac{p^+_u}{p^-_u}$.

In a linear framework, the elements of the transfer function, defined in Eq.~\eqref{eq:TransferMatrixFormula}, can be obtained by subjecting the system of ODEs in Eq.~\eqref{eq:SysOfODEs} to any two different conditions. For examples, $p^+_u = 0$, $p^-_u = 1$ and $p^+_u = 1$, $p^-_u = 0$ are chosen here. The acoustic response of the duct, expressed in terms of $\dfrac{p^+_d}{p^-_u}$ and $\dfrac{p^-_d}{p^-_u}$ is given by,

\begin{equation}
    \dfrac{p^{+}_{d}}{p^{-}_{u}} = T_{11} R_u + T_{12}; \quad \dfrac{p^{-}_{d}}{p^{-}_{u}} = T_{21} R_u + T_{22}; 
\end{equation}
where $T_{11}, T_{12}, T_{21}, T_{22}$ are the elements of the transfer matrix shown in Eq.~\eqref{eq:TransferMatrixFormula}. These expressions are finally used in the estimation of $\Delta$ using Eq.~\eqref{eq:Deltafandg}.

\section{Final analytical solution for conical ducts with linear stream-wise mean temperature gradient}\label{sec:appendixc}
The analytical solution for the acoustic field given by Eq.~\eqref{eq:FinalAnalyticalSolutionNonIsentropic} requires the evaluation of the terms under the integral sign. This can be achieved to reasonable accuracy by assuming the flow Mach number to have the following simplified dependence on area and mean temperature,
\begin{equation}\label{eq:FlowMachExpressionIsentropic_Appendix}
  \dfrac{M}{M_u} \approx \dfrac{A_u}{A} \sqrt{{\dfrac{{\overline{T}}}{\overline{T_u}}}}.
\end{equation}
This simplification is valid, as the analysis assumes low Mach number flows. It then follows that,
\begin{equation}\label{eq:P1plusminusNonIsentropicConical_Appendix}
        {\mathcal{P}^{\pm}}_{d} = {\left(\dfrac{{A}_{u}}{A_d}\right)}^{{1}/{2}}{\left(\dfrac{{\overline{T}}_{u}}{\overline{T}_{d}}\right)}^{{1}/{4}}\left[{\text{exp}\left(\dfrac{{M_d}^{2} - M_u^2}{2} \mp {\left(M_d-M_u\right)} + J_1^c \mp J_2^c \mp \text{i} I_0^c \pm \text{i} \left(  I_1^c + I_2^c + I_3^c + I_4^c \right)\right)}\right],
\end{equation}
where $J_1^c$ and $J_2^c$ (with $x_u^*$ = 0, $x_d^*$ = 1) are given as,
\begin{align}\label{eq:Terms3IntegrationAmplitudeNonIsentropicConical_Appendix}
J_1^c \: &= \: \displaystyle{\int _{ x_u^* }^{ x_d^* }} {\dfrac{\alpha^* \gamma M^2 }{2}\text{d}\breve{x}} \:\: =\:\: \dfrac{\gamma {M}_{u}^{2}}{12 a^*}\left[ (b^* + 3a^*) - \dfrac{4a^*b^* + b^* + 3a^*}{(1+a^*)^4} \right], \\
J_2^c \: &= \: \mp \displaystyle{\int _{ x_u^* }^{ x_d^* }} {\dfrac{\beta^* \gamma M}{2}\text{d}\breve{x} } \:\:=\:\: \dfrac{b^* M_u}{b^*-a^*}\left[ \dfrac{b^*}{\sqrt{a^*(b^*-a^*)}} \left( \text{tan}^{-1} \sqrt{\dfrac{a^*(1+b^*)}{b^*-a^*}} - \text{tan}^{-1} \sqrt{\dfrac{a^*}{b^*-a^*}}\right) + \dfrac{\sqrt{1+b^*}}{1+a^*} - 1 \right], 
\end{align} 
 and $I_0^c$ to $I_4^c$ correspond to the phase terms in Eq.~\eqref{eq:FinalAnalyticalSolutionNonIsentropic}.
\begin{equation}\label{eq:PhaseIntegrationNonIsentropicConical_Appendix}
\begin{aligned}
\int _{ {x_u^*} }^{x_d^*}{ \textrm{He}\left( \dfrac{\mp 1}{1 \pm M} \pm \dfrac{\Phi_{NI}^{\pm} }{2 \text{He}^2} \right) \text{d}\breve{x}}
      \approx  \overbrace{\int _{ x_u^* }^{ x_d^* } { \mp\text{He} \left(1 \mp M + M^2 \right) \text{d}\breve{x} }}^{\textbf{\normalsize{$I_o^c$}}} \pm  \overbrace{\int _{ x^*_u }^{ x_d^* } {\dfrac{1}{\text{He}}\left(\dfrac{\alpha^* \beta^*}{4} - \dfrac{{\beta^*}^2 }{32 } \mp \dfrac{3 {\alpha^*}^2 M}{4} \pm \dfrac{\alpha^* \beta^* M}{4}(3 - \gamma)\right) \text{d}\breve{x} }}^{\textbf{\normalsize{$I_1^c$} \: \small{to} \: \normalsize{$I_4^c$}}} .
\end{aligned}     
\end{equation} 
With $x_u^*$ = 0, $x_d^*$ = 1, these terms given by,
\begin{align}\label{eq:Terms1IntegrationPhaseNonIsentropicConical_Appendix}
    I_0^c =&\int _{ x_u^* }^{ x_d^* } { \text{He} \left(1 \mp M +M^2\right) \text{d}\breve{x} } = \text{He}_u \displaystyle{\int _{ x_u^* }^{ x_d^* }} {\left( \dfrac{1}{\sqrt{1+b^*}} \mp \dfrac{M_u}{\left(1+a^*\right)^2} + \dfrac{M_u^2 \sqrt{1+b^* })}{\left(1+a^* \right)^4}  \right) \text{d}\breve{x}}, \\ 
    &=\dfrac{2 \text{He}_u}{b^*} \left( \sqrt{1+b^*} -1 \right) \mp \dfrac{\text{He}_u M_u }{1 + a^*} \\ &+ \text{He}_u {M_u^2} \left[ \dfrac{{b^*} ^3 \tan^{-1}\left(\sqrt{\dfrac{a^* (1+ b^* x^*)}{b^* - a^*}}\right)}{8 {a^*}^{3/2} \left(b^* - a^*\right)^{5/2}} + \dfrac{\sqrt{1+b^* x^*} \: \left(b^*({a^* x^*} +3) -2a^*\right)  \left(b^*(3{a^* x^*} - 1) + 4a^*\right)}{24 a^* \left(b^*-a^*\right)^2 \left(1 + a^* x^* \right)^3}\right]^{{x^*_d\: = \:1}}_{{x^*_u\: = \:0}} 
\end{align}
\begin{align}\label{eq:Terms2IntegrationPhaseNonIsentropicConical_Appendix}
I_1^c \: &= \: \dfrac{1}{2} \displaystyle{\int _{ x_u^* }^{ x_d^* }} {\dfrac{\alpha \beta}{2 \textrm{He}}\text{d}\breve{x}}\:\: =\:\: \dfrac{a^*b^*}{\textrm{He}_u \sqrt{a^*(b^*-a^*)}} \left[\tan^{-1}\left(\sqrt{\dfrac{a^*(1+b^*)}{b^*-a^*}}\right) - \text{tan}^{-1}\left(\sqrt{\dfrac{a^*}{b^*-a^*}}\right) \right], \\
I_2^c \:&= \: \dfrac{-1}{16} \displaystyle{\int _{ x_u^* }^{ x_d^* }} {\dfrac{{\beta^*}^2}{2 \textrm{He}}\text{d}\breve{x}}\:\: =\:\: \dfrac{b^*}{16 \textrm{He}_u} \left(\dfrac{1}{\sqrt{1 + b^*}} - 1 \right), \\
I_3^c \:&= \:\mp \dfrac{3}{2} \displaystyle{\int _{ x_u^* }^{ x_d^* }} {\dfrac{{\alpha^*}^2 M}{2 \textrm{He}}\text{d}\breve{x}} \:\:= \:\:\mp \dfrac{M_u}{2 \textrm{He}_u} \left[\left(b^* + 2a^*\right) - \dfrac{3a^*b^* + b^* + 2a^*}{\left(1+a^*\right)^3}\right], \\
I_4^c \: &= \: \pm \dfrac{3-\gamma}{2} \displaystyle{\int _{ x_u^* }^{ x_d^* }} {\dfrac{\alpha^* \beta^* M}{2 \textrm{He}} \text{d}\breve{x}} \:\:= \:\:\mp \dfrac{(3 - \gamma) b^* M_u }{2\textrm{He}_u} \left[\dfrac{1}{(1+a^*)^2} - 1 \right]. 
\end{align}    
where He$_u$ is the Helmholtz number at the upstream end of the duct and $\breve{x}$ is a dummy variable. 

Thus, the expression for $\mathcal{P}^{\pm}_{d}$, for a conical duct sustaining non-isentropic mean flow is given by Eq.~\eqref{eq:P1plusminusNonIsentropicConical_Appendix}. As can be observed, this expression only depends on the mean flow parameters at the upstream $(x_u^*)$ and downstream $(x_u^*)$ ends of the duct.

\section{Effect of upstream boundary reflection coefficient}\label{sec:appendixd}
\begin{figure}[t]
\centering
\includegraphics[width=1\textwidth, height = 4.4in]{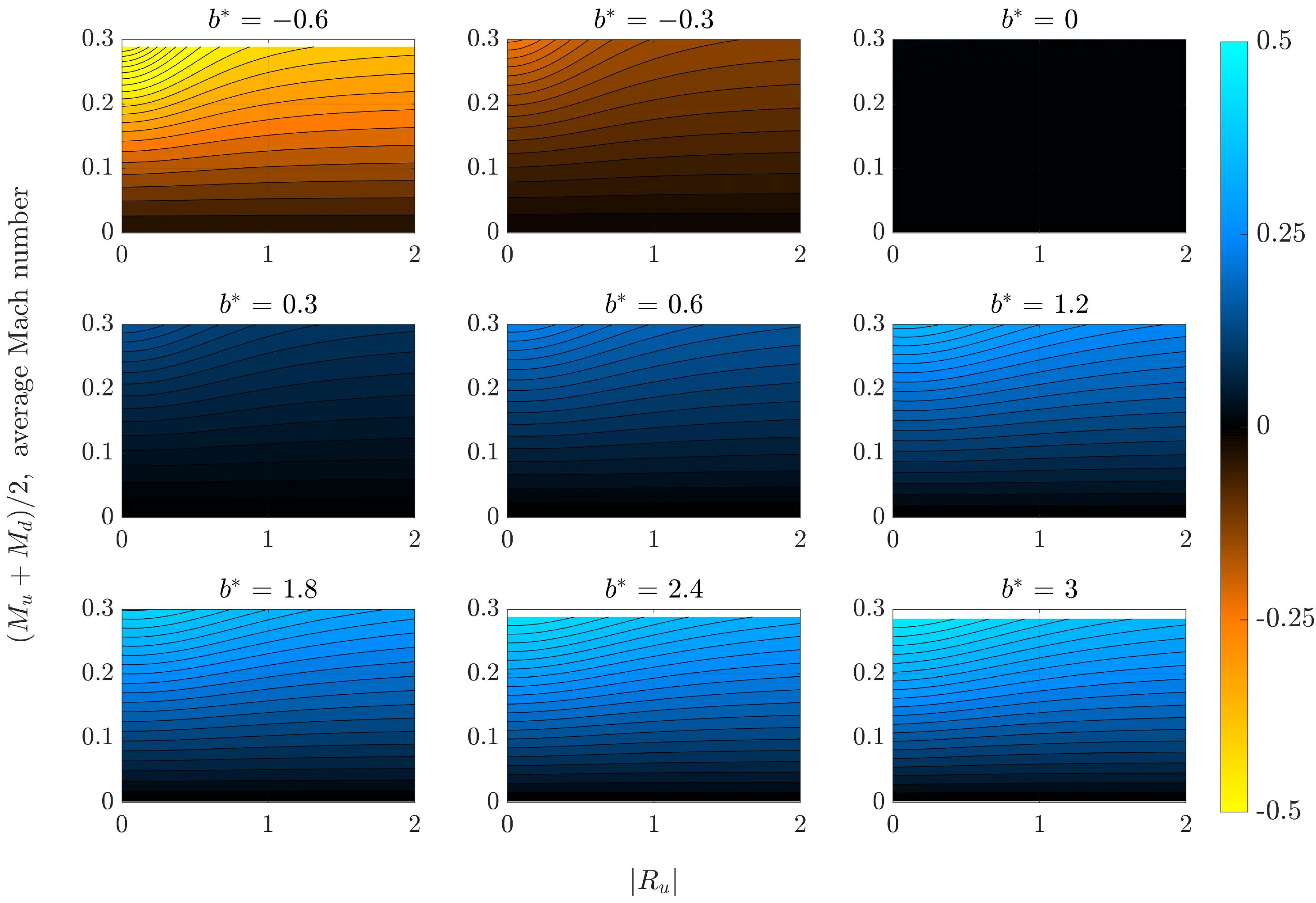}
\caption{Contours of $\Delta$ for $a^* =0.2$, He$_u$ = 1 as a function of the average flow Mach number, and absolute value of upstream boundary  reflection coefficient, for different values of mean temperature variation ($b^*$).}
\label{fig:Figure5}
\end{figure}

Fig.~\eqref{fig:Figure5} shows the contours of $\Delta$ as a function of absolute value of reflection coefficient $|R_u|$ and average flow Mach number $M_{avg}$, for a particular value of $a^*$ and different values of $b^*$, ranging from -0.6 to 3. It can be observed that for a negative temperature gradient ($b^* <0$), the minimum value of $\Delta$, implying maximum acoustic generation, occurs for an anechoic boundary at the upstream end ($|R_u|$ = 0). Similarly, when $b^* >0$, the maximum value of $\Delta$, corresponding to maximum acoustic absorption, again occurs for anechoic boundary at the upstream end ($|R_u|$ = 0). This suggests that the extrema of acoustic absorption coefficient $\Delta$ can be identified by considering only the case of an anechoic boundary condition at the upstream end.

\section{Optimisation search}\label{sec:appendixE}
The global extrema of $\Delta$ as a function of mean flow parameters ($a^*$, $b^*$ and $M_u$), are found using an optimisation search based on a genetic algorithm. The constraints applied to the parameters are as follows,
\begin{equation}
  -0.8 \leq a^{*}   \leq 2;\quad 
  -0.8 \leq b^{*}   \leq 5; \quad
   0   \leq {M}_{u} \leq 0.3.
\end{equation}
Although the theoretical limits for the values of $a^*$ and $b^*$ range from -1 to $\infty$, these area and temperature profiles are practically infeasible. For example, $a^* > 2$ indicates a diverging duct with significant area change and hence prone to flow separation. Thus, the non-dimensional area variation $a^*$ is limited by nominal levels of flow acceleration for converging ducts and flow separation for diverging ducts. The non-dimensional mean temperature variation $b^*$ is limited by the combustion process and the practical temperatures gradients encountered. The value of $M_u$ is limited by the low Mach number flow assumption, typical for combustor flows. The flow conditions for maximum acoustic absorption/generation are presented in Table~\ref{table:1}, along with their corresponding values.

\begin{table*}[ht]
\begin{center}
\caption{The flow conditions and corresponding extrema for $\left|\Delta\right|$. \label{table:1}}
 \begin{tabular}{||c c||} 
 \hline
  {\textbf{Acoustic generation}} & {\textbf{Acoustic absorption}} \\ 
 [0.5ex] 
 \hline\hline
  $a^* = -0.373$ & $a^* = -0.72$ \\ 
  \hline
  $b^{*} = -0.8$ & $b^* = 5$ \\
  \hline
  ${M}_{{u}} = 0.25$& $M_u = 0.262$ \\ [1ex]
  \hline\hline
  $\Delta_{\textrm{min}}\: =\: -1.335$ & $\Delta_{\textrm{max}}\: =\: 0.611$ \\ 
 [0.5ex] 
 \hline
\end{tabular}
\end{center}
\end{table*}
As can be seen from Table~\ref{table:1}, maximum acoustic generation ($\Delta$ = -1.335) occurs when the non-dimensional mean temperature variation $b^*$ assumes the lowest possible negative value constrained by the limit. Similarly, the maximum acoustic absorption ($\Delta$ = 0.611) corresponds to $b^*=5$, constrained by the maximum positive value set by the limits. 

\section{Simplified mathematical analysis}\label{sec:appendixF}
The effects of the mean flow, frequency and upstream boundary reflection coefficient on the acoustic absorption coefficient are further investigated in this section using a simplified mathematical analysis, wherein low subsonic flow Mach numbers are considered (only terms up to O$(M)$ are retained). Firstly, ducts sustaining isentropic mean flow ($\overline{\dot{Q}} = 0$) with an anechoic upstream boundary ($R_u = 0$) are considered to understand the effects of frequency (He$_u$) and area variation ($\alpha^*$). This is followed by analysing ducts with non-isentropic mean flows and a non-anechoic upstream boundary ($R_u \neq 0$).

\subsection{ Varying area ducts with isentropic mean flow}\label{ssec:AreaEffect_Appendix}
For a duct with isentropic mean flow ($\overline{\dot{Q}} = 0$) with negligible flow Mach numbers, the variation in temperature and pressure can be obtained from  Eqs.~\eqref{eq:HeatTransferDefinition} and~\eqref{eq:gradpNonIsentropic}:
\begin{equation}
 \dfrac{1}{\overline{T}}\dfrac{\text{d}\overline{T}}{\text{d}x^*} = \dfrac{-\alpha^* {M}^{2}\left(\gamma-1\right)}{1 - \gamma{M}^{2}} \approx 0 \quad \Rightarrow \; \Xi \approx 1; \; \textrm{also,} \quad
  \dfrac{1}{\overline{p}}\dfrac{\text{d}\overline{p}}{\text{d}x^*} = \dfrac{-\left(\beta^*-\alpha^*\right) \gamma {M}^{2}}{1 - \gamma{M}^{2}} \approx 0 \; \Rightarrow \; \xi \approx 1,.
\end{equation}
and therefore $\Delta$ is given by,
\begin{align}\label{eq:simpleDeltaarea}
\Delta_{|R_u|=0} =  1 -  \dfrac{{\left(1 - 2 M_u\right)\theta \:  +  {{{\left(1 + 2M_d\right)}}\left|{{p}^+_d}/{{p}^-_u}\right|}^{2}}} {{{\left(1 - 2M_d\right)}}{{\left|{{p}^-_d}/{{p}^-_u}\right|}}^{2}}.
\end{align}

Using Eq.~\eqref{eq:OutletToInlet}, we derive explicit expressions for $\left|{{p}^+_d}/{{p}^-_u}\right|^2$, $\left|{{p}^-_d}/{{p}^-_u}\right|^2$ that depend on the terms ${\mathcal{P}_{d,\:i}^{\pm}}$, ${\mathcal{B}_{u,\:d;\:i}^{\pm}}$. These terms $\mathcal{P}^{\pm}_{d}$ and $\mathcal{B}^{\pm}_{u,\:d;\:i}$ are given in \citep{Yeddula_JSV_2020} for the case of an isentropic mean flow and on defining $\delta_{u,\:d} = \dfrac{\alpha^*_{u,\:d}}{\text{He}_{u,\:d}}$, neglecting terms of order O$(M^2)$, O$(\delta^2)$ and higher, they simplify to,
\begin{align}\label{eq:SimpleP+P-B+area}
\mathcal{P}^{\pm}_{d,\:i} = \sqrt{\theta} \:\textrm{exp}\left(\pm S_1\mp \textrm{i}\:S_2\right), \;
\mathcal{B}^{\pm}_{u,\:d;\:i} = 1 \mp \dfrac{\textrm{i}\delta_{u,\:d}}{2} + {\textrm{i}\delta_{u,\:d} M_{u,\:d}},\; \textrm{where}\; S_1 = (M_u - M_d), \; S_2 =  \int _{ {x}_{u} }^{ x_d }{ \left( {1 \mp M}\right)\text{d}x^*}, 
\end{align}
and the corresponding expressions for $\left|{{p}^+_d}/{{p}^-_u}\right|^2$, $\left|{{p}^-_d}/{{p}^-_u}\right|^2$ read,
\begin{equation}\label{eq:finalp+dp-uarea}
\left|\dfrac{{p}^{+}_d}{{p}^{-}_u}\right|^2 = \dfrac{\theta}{4} \left[ \delta_u^2 \: \Gamma_u^2 \: \textrm{exp}\left(2S_1\right) + \delta_d^2 \: \Gamma_d^2 \: \textrm{exp}\left(-2S_1\right) - 2\left(\delta_u \Gamma_u\right)\: \left(\delta_d \Gamma_d\right) \textrm{cos}\left(2h\right) \right],
\end{equation}
and,
\begin{equation}\label{eq:finalp-dp-uarea}
 \left|\dfrac{{p}^{-}_d}{{p}^{-}_u}\right|^2 = \theta \: \textrm{exp}\left(-2S_1\right) \Bigg[ 1 +  \left({\dfrac{\delta_u \Gamma_u - \delta_d \Gamma_d}{2}}\right)^2 \Bigg].
\end{equation}
where, $\Gamma_{u,\:d} = 1+2M_{u,\:d}$, $h = \textrm{He}_u\: S_2$.
Substituting Eqs.~\eqref{eq:finalp+dp-uarea} and~\eqref{eq:finalp-dp-uarea} in  Eq.~\eqref{eq:simpleDeltaarea}, neglecting terms of order O$(\delta^2)$ and using $M_d = M_u \theta$ (Eq.~\eqref{eq:FlowMachExpressionIsentropic_Appendix}), we arrive at an expression for $\Delta$ that depends only on the mean flow and its gradients,
\begin{equation}\label{eq:DeltaMeanArea}
\Delta_{|R_u|=0} =  1 - \dfrac{\left(1-2M_u\right)} { \left(1-2{M_u \:\theta}\right)\: \textrm{exp}\left(-2 M_u \left(1 - \theta\right)\right)}
\end{equation}

As Eq.~\eqref{eq:DeltaMeanArea} implies that the acoustic absorption coefficient depends only on the mean flow, any dependence on frequency (not captured by Eq.~\eqref{eq:DeltaMeanArea}) must be weak. This dependence on frequency is negligible for small values of $\delta$ - a prerequisite for using the WKB solution. Thus the mathematical analysis again confirms that the effect of frequency (He$_u$) on the acoustic absorption coefficient ($\Delta$) is negligible as long the frequency criteria (Eqs.~\eqref{eq:FrequencyLimitLower},~\eqref{eq:FrequencyLimitUpper}) are satisfied. 

On further expanding the exponential function in Eq.~\eqref{eq:DeltaMeanArea} using Taylor's series and neglecting terms of order O$(M^2)$ and higher we obtain, 
\begin{equation}\label{eq:DeltaMeanSimplearea}
\Delta_{|R_u|=0} =  1 -  \dfrac{\left(1-2M_u\right)} { \left(1-2{M_u \:\theta}\right)\: \left(1 - 2 M_u \left(1 - \theta\right)\right)} = 1 - \dfrac{1-2M_u}{1-2M_u} = 0;
\end{equation}

It can be observed that for ducts of arbitrary profile, the acoustic absorption coefficient equals zero for the case of an isentropic mean flow. However, at low flow Mach numbers, both isentropic and isothermal mean flows are associated with negligible and zero temperature variations respectively,
\begin{equation}
    \beta^*_i = \dfrac{1}{\overline{T}}\dfrac{\text{d}\overline{T}}{\text{d}x^*} = \dfrac{-\alpha^* {M}^{2}\left(\gamma-1\right)}{1 - \gamma{M}^{2}} \approx 0, \quad \text{while} \quad \beta^*_{\overline{T} = c} = 0.
\end{equation}

Hence, the acoustic absorption coefficient $\Delta$ dependence on the temperature profile is more pronounced, since the main source of mean flow non-isentropicity is the heat transfer (or the temperature gradient), leading to the absorption or generation of acoustic energy in duct flows.

\subsection{Straight ducts with non-isentropic mean flow}
As seen in \ref{ssec:AreaEffect_Appendix} the effect of the area variation ($\alpha^*$) and frequency (He$_u$) on $\Delta$ are negligible for ducts sustaining low flow Mach numbers. Hence, this section analyses the effects of the absolute value of the reflection coefficient $|R_u|$ and the temperature variation $b^*$ on the acoustic absorption coefficient $\Delta$, assuming straight ducts to keep the procedure simple. A similar procedure as outlined in \ref{ssec:AreaEffect_Appendix} is used in deriving an simplified expression for $\Delta$ by neglecting O$(M^2)$ and O$(\delta^{2})$, which reads, 
\begin{equation}\label{eq:simpleDeltaRu}
\Delta \approx  1 -  \dfrac{{\left(1 - 2 M_u\right) \:  +  {{{\left(1 + 2M_d\right)}}\:\textrm{exp}\left(2S_3\right)\left|R_u\right|}^{2}}}  {\left(1+2M_u\right) {\left|R_u\right|}^2 + {{\left(1 - 2M_d\right)}}\:\textrm{exp}\:\left(-2S_3\right) }, \quad \text{where,}
\quad S_3 = -M_u \left(\dfrac{1}{\Xi} + \gamma\sqrt{1+b^*} - \left(\gamma+1\right)\right).
\end{equation}
Expanding the exponential series by Taylor's expansion up to O$(M)$ in Eq.~\eqref{eq:simpleDeltaRu} gives,
\begin{equation}\label{eq:DeltaMeanRu}
\Delta \approx  \dfrac{\overbrace{2M_u \gamma \left(\sqrt{1+b^*} - 1\right) {\left|R_u\right|}^{2}}^{F_1}+ 2M_u \gamma \left(\sqrt{1+b^*} - 1\right)}  {\underbrace{\left(1+2M_u\right) {\left|R_u\right|}^2}_{F_2} + \left(1 + 2M_u \left(\gamma\sqrt{1+b^*} - \left(\gamma+1\right)\right)\right)}.
\end{equation}
It can be shown that maxima of the acoustic absoprtion coefficient $|\Delta|$ occur for anechoic upstream boundary by considering the cases of positive ($b^*_+$) and negative ($b^*_-$) values of temperature variation ($b^*$) separately.

In the above equation for $\Delta$ (Eq.~\eqref{eq:DeltaMeanRu}), $F_1$ and $F_2$ are the terms that depend upon the reflection coefficient at the upstream end ($R_u$). When the temperature parameter $b^*$ is positive ($b^*_+ >0$), the condition $\Delta_{|R_u|=0}$ \textgreater $\Delta$ requires,
\begin{equation}\label{eq:b^*+}
    b^*_+ \; \textless \; \dfrac{1+ 4M_u (1+\gamma) + 4 M_u^2 \left(1+ 2\gamma\right)}{4 M_u^2 \gamma^2}.
\end{equation}
The above inequality is always satisfied owing to the $M_u^2$ term in the denominator in the limit of low flow Mach numbers considered.
Similarly for a negative value of the temperature parameter $b^*_- = -\left|b^*\right|$, the condition $\left|\Delta_{|R_u=0|}\right|$  \textgreater $\left|\Delta\right|$ takes the form,
\begin{equation}\label{eq:b*-}
 \left|b^*\right| > \dfrac{- \left(1+ 4 M_u \left(1-\gamma\right)+ 4 M_u^2 \left(1-2\gamma\right) \right)}{4 M_u^2 \gamma^2} \quad\Rightarrow \quad M_u < \dfrac{1}{4\left(\gamma-1\right)}.
\end{equation}

For the considered values of low flow Mach numbers, this inequality is again always satisfied. For example, if $\gamma = 1.4$, inequality in Eq.~\eqref{eq:b*-} takes the form $M_u \:\textless \:0.625$.
Therefore, it can be generalised for both the cases of positive and negative values of $b^*$ that $\left|\Delta_{|R_u|=0}\right|  > \left|\Delta_{|R_u|\neq0}\right|$ in the limit of low subsonic flow Mach numbers and sufficiently high frequencies. This behaviour is also confirmed using numerical estimations for $\Delta$, shown in \ref{sec:appendixd}. 

It can also be noticed from the equation Eq.~\eqref{eq:DeltaMeanRu} that any positive value of the temperature profile ($b^*_+$) ensures $\Delta \: \textgreater \: 0$, while negative values ($b^*_-$) give $\Delta \textless 0$. This again showing that positive and negative values of the temperature gradient, $b^*$, correspond to acoustic energy absorption and generation respectively.

\end{document}